\documentclass[aps,twocolumn,groupedaddress]{revtex4}
\usepackage{epsfig}
\usepackage{graphicx}
\usepackage[T1]{fontenc}
\usepackage{ae}
\usepackage[latin1]{inputenc}
\usepackage{amssymb,amsbsy,amsmath}
\usepackage{bbm}

\begin{document}



\title[Theory of ground states II]
{Theory of ground states for classical Heisenberg spin
systems II}

\author{Heinz-J\"urgen Schmidt$^1$
\footnote[1]{Correspondence should be addressed to hschmidt@uos.de}
}
\address{$^1$Universit\"at Osnabr\"uck, Fachbereich Physik,
Barbarastr. 7, D - 49069 Osnabr\"uck, Germany}


\begin{abstract}
We apply the theory of ground states for classical, finite, Heisenberg spin systems previously published to
a couple of spin systems that can be considered as finite models $K_{12},\,K_{15}$ and $K_{18}$ of the AF Kagome lattice.
The model $K_{12}$ is isomorphic to the cuboctahedron.
In particular, we find three-dimensional ground states that cannot be viewed as resulting from
the well-known independent rotation of subsets of spin vectors. For a couple of ground states with translational symmetry
we calculate the corresponding wave numbers.
Finally we study the model $K_{12w}$ without boundary conditions which exhibits new phenomena as, e.~g., two-dimensional families
of three-dimensional ground states.
\end{abstract}

\maketitle

\section{Introduction}\label{sec:I}
The theory outlined in \cite{S17} and \cite{S17b} makes it possible to calculate, in principle, all classical ground states of a finite Heisenberg system.
There are two restrictions to this general claim: (1) all ground states calculated in this way may be $M$-dimensional with $M>3$ and hence un-physical,
and (2) there are practical restrictions to the calculations if the number $N$ of spins is too large. The first restriction is supposed to appear
rarely in practice. In order to assess the second restriction one has to evaluate examples that are more complex than those considered in \cite{S17}.
As such an example we investigate in this paper the Kagome lattice that has been the subject of a vast number of articles. Here we only mention
a small selection of papers also concerned  with classical ground states of the Kagome lattice, see \cite{S92}--\cite{BZM17}.

The paper is organized as follows. In section \ref{sec:D} we recapitulate some results of \cite{S17} in a form suited for the present purpose,
confining ourselves to the case of an ``undressed ${\mathbbm J}$-matrix". The Kagome lattice is shortly described in section \ref{sec:K} and
the theory outlined in \cite{S17} is applied to three finite Kagome models in the subsections \ref{sec:K12}, \ref{sec:K15}, and \ref{sec:K18}.
In order to illustrate the influence of periodic boundary conditions we consider also a model with $N=12$ spins without boundary conditions in
section \ref{sec:K12WB}. It is necessary to introduce the ``dressed ${\mathbbm J}$-matrix" in this case and to slightly extend the theory presented in section \ref{sec:D}. We close with a summary and outlook.

\section{General definitions and results}\label{sec:D}
 A configuration
${\mathbf s}_\mu,\;\mu=1,\ldots,N$ of $M$-dimensional spin vectors can be represented by its ``Gram matrix" $G$ with entries
\begin{equation}\label{D1}
  G_{\mu\nu}={\mathbf s}_\mu\cdot{\mathbf s}_\nu,\;\mu,\nu=1,\ldots,N.
\end{equation}
Two spin configurations have the same Gram matrix iff they are equivalent w.~r.~t.~a global rotation/reflection $R\in O(M)$.
Let
\begin{equation}\label{D2}
 H=\sum_{\mu,\nu=1}^N J_{\mu\nu}\,{\mathbf s}_\mu\cdot{\mathbf s}_\nu=\mbox{ Tr} \left({\mathbbm J}\,G\right)
\end{equation}
be the Hamiltonian of the spin system where ${\mathbbm J}$ denotes the symmetric  $N\times N$-matrix with entries $J_{\mu\nu}$.
Let $E_{min}$ denote the global minimum of $H$. Then all Gram matrices of ground states, defined by $\mbox{ Tr} \left({\mathbbm J}\,G\right)=E_{min}$,
are of the form
\begin{equation}\label{D3}
G=W\,\Delta\,W^\top\;,
\end{equation}
where $W$ is some $N\times M$-matrix the columns of which span the eigenspace of ${\mathbbm J}$ corresponding to its lowest eigenvalue
$j_{min}$ and $\Delta$ is some positively semi-definite $M\times M$-matrix that is a solution of the ``additional degeneracy equation" (ADE)
\begin{equation}\label{D4}
\left(W\,\Delta\,W^\top\right)_{\mu\mu}=1\mbox{    for all   }\mu=1,\ldots,N.
\end{equation}
The convex set of solutions $\Delta\ge 0$ of the ADE is denoted by ${\mathcal S}_{ADE}$. We stress that these results are only valid for a certain
subclass of spin systems including finite models of the Kagome lattice with periodic boundary conditions.
For the general theory involving ``dressed ${\mathbbm J}$-matrices"
we refer the reader to \cite{S17} and to the corresponding remarks of section \ref{sec:K12WB} where the reduced Kagome model $K_{10}$ without boundary conditions is treated.

Permutations $\pi\in{\mathcal S}_N$ of the $N$ spin sites are represented by $N\times N$-matrices $\Pi\in{\sf S}_N$ by correspondingly permuting
the standard basis of ${\mathbbm R}^N$. Let ${\sf Gr}$ denote the group of ``symmetries" consisting of all $\Pi\in{\sf S}_N$ that commute
with ${\mathbb J}$. Let ${\sf G}\subset {\sf Gr}$ be some subgroup of symmetries, then a state ${\mathbf s}$
is called ${\sf G}$-symmetric if its Gram matrix $G$ commutes with all $\Pi\in{\sf G}$. It follows that in this case $\Pi\,{\mathbf s}={\mathbf s}\,R$
for some $R\in O(M)$. The existence of symmetric ground states can be easily proven, see \cite{S17}. We denote by ${\mathcal S}_{ADE}^{sym}$
the set of solutions of the ADE that lead to ${\sf G}$-symmetric ground states. It is a convex subset of ${\mathcal S}_{ADE}$.
If the set ${\mathcal S}_{ADE}$ is too large to be analyzed in detail we thus may confine ourselves to ${\mathcal S}_{ADE}^{sym}$.

A particular case is an Abelian subgroup ${\sf T}\subset {\sf Gr}$ of ``translations" that can be defined for certain finite models of spin lattices.
In this case the eigenvalues of $R\in O(M)$ corresponding to a ${\sf T}$-symmetric ground state ${\mathbf s}$ in the way described above are related to the ``wave-numbers" $q$ of ${\mathbf s}$, see the examples below.

\section{The Kagome lattice}\label{sec:K}
\begin{figure}[ht]
  \centering
    \includegraphics[width=1.0\linewidth]{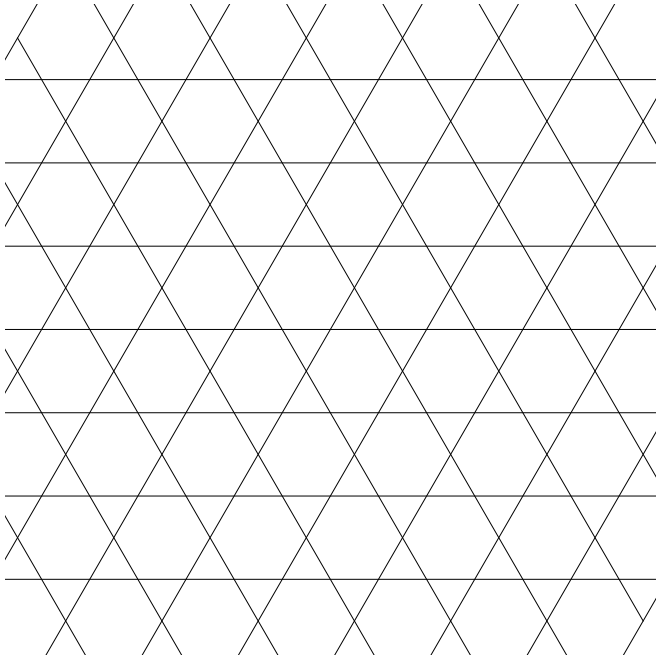}
  \caption
  {A detail of the infinite Kagome lattice.
  }
  \label{FIGKAGO}
\end{figure}

The Kagome lattice is a plane, infinite lattice consisting of triangles and hexagons, see Figure \ref{FIGKAGO}. Its vertices can be defined as
vectors of the form
\begin{equation}\label{K1}
[\mu,\nu]\equiv \mu \, {1\choose 0} +\nu\,{1/2\choose\sqrt{3}/2}
\;,
\end{equation}
such that $(\mu,\,\nu)$ runs through all pairs of integers except those where both, $\mu$ and $\nu$, are odd.
Let us denote the set of such integer pairs as ${\mathcal K}$ which is also used to denote the Kagome lattice.
The Kagome lattice has been used
as a model for an infinite spin system by taking the vertices as spin sites and introducing a, say, uniform AF Heisenberg coupling between adjacent
spin sites. The detailed properties of the Kagome spin lattice have to be defined in terms of an appropriate thermodynamic limit, which need not be considered here.
The only property of the classical Kagome spin lattice we need is that any infinite spin configuration ${\mathbf s}_{\mu\nu},\;(\mu,\nu)\in{\mathcal K}$
that minimizes the energy of each local triangle, i.~e.~the spin vectors of each triangle forming mutual angles of $120^\circ$, will also be a ground state
of the Kagome lattice (in a sense to be made precise in the definition of the thermodynamic limit). It is obvious that there exists an infinite number of co-planar
spin configurations that are ground states in the above sense. Moreover, it is well-known that certain families of $3$-dimensional ground states
can be constructed by independent rotations of spin vectors within certain subsets of ${\mathcal K}$, see, e.~g., \cite{S92}. In this paper we will
describe further $3$-dimensional ground states of the Kagome lattice.

Since the theory of ground states outlined in \cite{S17} is limited to finite spin systems we will have to consider ``finite models" of the Kagome
lattice. This is a widely established practice in the field of numerical investigations of the Kagome spin lattice. Mathematically a finite model
can be represented by an equivalence relation $\sim$ on ${\mathcal K}$ such that the set of equivalence classes ${\mathcal K}/\sim$ is finite.
Two equivalence classes are defined to be adjacent iff there exist representatives that are adjacent spin sites in the Kagome lattice.
If  ${\mathcal K}/\sim$ is represented by a finite set of spin sites of ${\mathcal K}$ this definition entails the introduction of ``periodic
boundary conditions".

Recall that the group ${\mathcal T}$ of translations of the Kagome lattice is generated by the two maps $S:[\mu,\nu]\mapsto[\mu+2,\nu]$
and $T:[\mu,\nu]\mapsto [\mu,\nu+2]$. The set of spin sites ${\mathcal K}$ is a disjoint union of three orbits of ${\mathcal T}$.
The three spin sites $[0,0],\,[0,1]$ and $[1,0]$ are representatives of these orbits and accordingly said to form a ``primitive unit cell".
The mentioned equivalence relation on ${\mathcal K}$ can be generated by a suitable subgroup ${\mathcal T}_0$ of ${\mathcal T}$
such that two spin sites are equivalent iff they can be connected by a translation $L\in {\mathcal T}_0$. The corresponding set
of equivalence classes of spin sites will also be denoted by ${\mathcal K}/{\mathcal T}_0$

Let us consider a simple example, where the subgroup ${\mathcal T}_0$ of ${\mathcal T}$ consists of all ``even" translations
$L:[\mu,\nu]\mapsto[\mu+4 k,\nu+4 \ell],\; k,\ell\in {\mathbbm Z}$. Then a representative set of ${\mathcal K}/{\mathcal T}_0$ is
given by the $12$ spin sites $(\mu,\nu)\in{\mathcal K}$ with $0\le \mu,\nu\le 3$, see Figure \ref{FIGK1}.
There are $4^2=16$ such pairs $(\mu,\nu)$, but $(1,1), (1,3), (3,1)$ and $(3,3)$ have to be excluded since both numbers are odd.
The resulting finite Kagome model will be called $K_{12}$.

For the two other models $K_{15}$ and $K_{18}$ used in this paper the corresponding subgroups are defined as follows.
For the first case let ${\mathcal T}_0$ be generated by the maps $T\,S^2$ and $T^2\,S^{-1}$. Equivalently, ${\mathcal T}_0$ can be defined
as the subgroup of maps $T^m\,S^n$ such that $m+2n$ and $2m-n$ are integer multiples of $5$. It follows that $\left|{\mathcal K}/{\mathcal T}_0\right|=15$,
see Figure \ref{FIGK15}.

For the second case let ${\mathcal T}_0$ be generated by the maps $S^3$ and $T^2\,S^{-1}$. Equivalently, ${\mathcal T}_0$ can be defined
as the subgroup of maps $T^m\,S^n$ such that $m+2n$ is an integer multiple of $6$ and $m$ is even.
It follows that $\left|{\mathcal K}/{\mathcal T}_0\right|=18$,
see Figure \ref{FIGK5}.

For any finite model $K_n$ of the Kagome lattice it follows that any state ${\mathbf s}^{(n)}$ of $K_n$ can be periodically extended to a state
${\mathbf s}$ of  ${\mathcal K}$. Moreover, if ${\mathbf s}^{(n)}$ is a ground state of  $K_n$  then also ${\mathbf s}$ will be ground state of  ${\mathcal K}$
because ${\mathbf s}$ minimized each triangle Hamiltonian. This is a peculiar property of the Kagome lattice that will not hold for general lattices.

\subsection{The $N=12$ Kagome model}\label{sec:K12}
\begin{figure}[ht]
  \centering
    \includegraphics[width=1.0\linewidth]{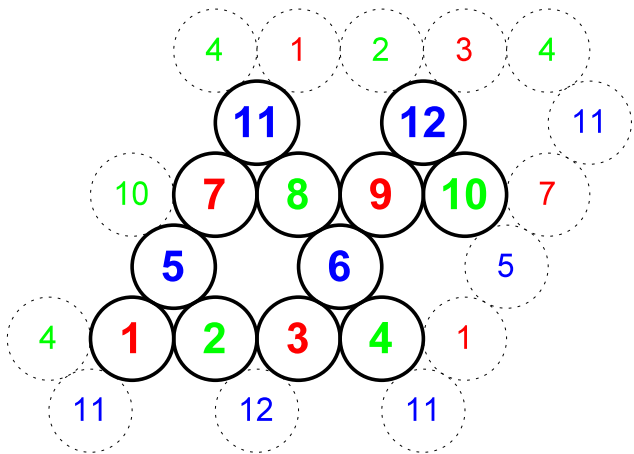}
  \caption
  {Representation of a finite $N=12$ model of the AF Kagome lattice. A co-planar ground state ${\mathbf a}_{12}$ with three spin directions forming mutual angles of
  $120^\circ$ is indicated by a coloring of the spin sites with the colors red, green, blue.
  }
  \label{FIGK1}
\end{figure}

\begin{figure}[ht]
  \centering
    \includegraphics[width=1.0\linewidth]{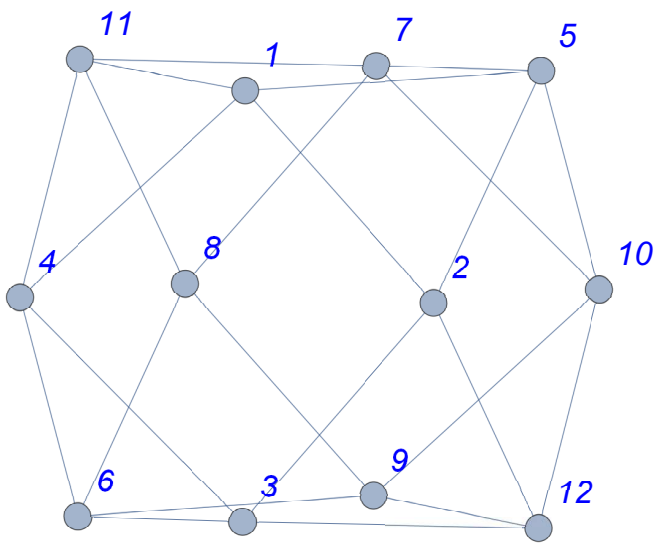}
  \caption
  {Representation of the graph underlying the $N=12$ model of the AF Kagome lattice. It is obvious that this is the graph of a cuboctahedron
  that results from joining the $12$ midpoints of the edges of a cube.
  }
  \label{FIGK1a}
\end{figure}
A finite model $K_{12}$ of the AF Kagome lattice with periodic boundary conditions is shown in Figure \ref{FIGK1}.
It is obvious from the representation of Figure \ref{FIGK1a} that  $K_{12}$ is graph-theoretically isomorphic to the cuboctahedron.
Hence the results of this subsection also apply to the AF cuboctahedron.
As noted above, any state where the neighboring spin vectors form mutual angles of $120^\circ$ will be a ground state
of $K_{12}$ and can be periodically extended to a ground state of the Kagome lattice. One example is the co-planar ground state ${\mathbf a}_{12}$ indicated in
Figure \ref{FIGK1} by coloring the spin sites with the colors red, green, blue. It can be generated by periodic extension of a coloring of the
primitive unit cell of spin sites with the numbers, say, $1,\,2$ and $5$.
One observes that the spin vectors of the rows $1,2,3,4$ and $7,8,9,10$
can be independently rotated about the ``blue" spin axis without loosing the ground state property. This also holds generally for the Kagome lattice
in the sense that every second row can be independently rotated.
For the model $K_{12}$ this independent rotation yields a $1$-parameter family of $3$-dimensional ground states.
After a rotation of $180^\circ$ we again obtain a co-planar
ground state, called $b_{12}$, that results from $a_{12}$ by, say, interchanging the colors red and green in the row $7,8,9,10$.\\
Similarly, the above construction of ground states can be repeated by independent rotation of the lines $1,5,7,11$ and $3,6,9,12$ about the
green spin axis, or,  by independent rotation of the lines $4,6,8,11$ and $12,2,5,10$ about the red spin axis. Note that the latter line is
formed by periodic continuation. By this procedure we obtain two further $1$-parameter families of $3$-dimensional ground states
joining the co-planar ground state $a_{12}$ with other co-planar ground states $c_{12}$ and $d_{12}$.
All these ground states are well-known, see, e.~g., \cite{S92}, or \cite{S10} for the analogous considerations of ground states of the cuboctahedron.\\

\begin{figure}[ht]
  \centering
    \includegraphics[width=1.0\linewidth]{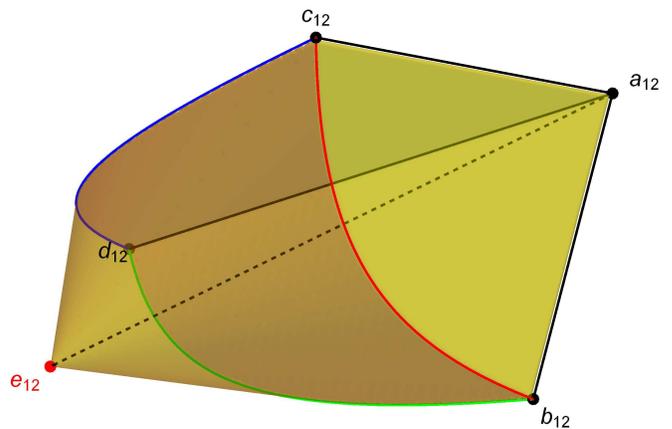}
  \caption
  {The convex set ${\mathcal S}_{12} $ the points of which correspond to $O(M)$-equivalence classes of ground states of $K_{12}$.
  The four extremal points $a_{12}$, $b_{12}$, $c_{12}$, and $d_{12}$ (black color) correspond to the co-planar ground states described above;
  the fifth extremal point $e_{12}$ (red color) that
  is the vertex of the cone ${\mathcal C}$ corresponds to an isolated $3$-dimensional ground state.
  The three black lines joining $a_{12}$ with $b_{12},c_{12}$ and $d_{12}$ correspond
  to $1$-parameter families of $3$-dimensional ground states generated by independent rotations of subsets of spins. The three
  colored  hyperbolic curves
  joining $b_{12}$ with $c_{12}$, $b_{12}$ with $d_{12}$, and $b_{12}$ with $d_{12}$ are additional $1$-parameter families of $3$-dimensional ground states.
  Finally, we have indicated by a black dashed line joining $a_{12}$ and $e_{12}$ the set of points corresponding to ${\sf Gr}_h$-symmetric ground states.
  }
  \label{FIGK2}
\end{figure}

Now we will apply the theory outlined in \cite{S17} in order to obtain further $3$-dimensional ground states not available by the above procedure.
The model $K_{12}$ has the ${\mathbbm J}$-matrix
\begin{equation}\label{K121}
 {\mathbbm J}=
 \left(
\begin{array}{cccccccccccc}
 0 & 1 & 0 & 1 & 1 & 0 & 0 & 0 & 0 & 0 & 1 & 0 \\
 1 & 0 & 1 & 0 & 1 & 0 & 0 & 0 & 0 & 0 & 0 & 1 \\
 0 & 1 & 0 & 1 & 0 & 1 & 0 & 0 & 0 & 0 & 0 & 1 \\
 1 & 0 & 1 & 0 & 0 & 1 & 0 & 0 & 0 & 0 & 1 & 0 \\
 1 & 1 & 0 & 0 & 0 & 0 & 1 & 0 & 0 & 1 & 0 & 0 \\
 0 & 0 & 1 & 1 & 0 & 0 & 0 & 1 & 1 & 0 & 0 & 0 \\
 0 & 0 & 0 & 0 & 1 & 0 & 0 & 1 & 0 & 1 & 1 & 0 \\
 0 & 0 & 0 & 0 & 0 & 1 & 1 & 0 & 1 & 0 & 1 & 0 \\
 0 & 0 & 0 & 0 & 0 & 1 & 0 & 1 & 0 & 1 & 0 & 1 \\
 0 & 0 & 0 & 0 & 1 & 0 & 1 & 0 & 1 & 0 & 0 & 1 \\
 1 & 0 & 0 & 1 & 0 & 0 & 1 & 1 & 0 & 0 & 0 & 0 \\
 0 & 1 & 1 & 0 & 0 & 0 & 0 & 0 & 1 & 1 & 0 & 0 \\
\end{array}
\right).
\end{equation}
According to the equivalence of all spin sites we need not consider the ``dressed" ${\mathbbm J}$-matrix and
all ground states of $K_{12}$ live on the $5$-dimensional eigenspace of ${\mathbbm J}$ corresponding to its lowest eigenvalue $j_{min}=-2$.
This eigenspace is spanned by the $5$ columns of the matrix
\begin{equation}\label{K122}
W=\left(
\begin{array}{ccccc}
 0 & -1 & 0 & 0 & -1 \\
 0 & 0 & 1 & -1 & 1 \\
 -1 & 0 & -1 & 1 & -1 \\
 0 & 0 & 0 & 0 & 1 \\
 0 & 1 & -1 & 1 & 0 \\
 1 & 0 & 1 & -1 & 0 \\
 0 & -1 & 0 & -1 & 0 \\
 0 & 0 & 0 & 1 & 0 \\
 -1 & 0 & -1 & 0 & 0 \\
 0 & 0 & 1 & 0 & 0 \\
 0 & 1 & 0 & 0 & 0 \\
 1 & 0 & 0 & 0 & 0 \\
\end{array}
\right).
\end{equation}
The corresponding general solution $\Delta$ of the ADE (\ref{D4}) depends on three real parameters called $x,y,z$:
\begin{equation}\label{K123}
 \Delta=
 \left(
\begin{array}{ccccc}
 1 & x & -\frac{1}{2} & y & y \\
 x & 1 & y & -\frac{1}{2} & -\frac{1}{2} \\
 -\frac{1}{2} & y & 1 & \frac{1}{2}-y & z \\
 y & -\frac{1}{2} & \frac{1}{2}-y & 1 & y+z+\frac{1}{2} \\
 y & -\frac{1}{2} & z & y+z+\frac{1}{2} & 1 \\
\end{array}
\right).
\end{equation}
Let us denote by ${\mathcal S}_{12}$ the convex domain in the $x,y,z$-space such that $\Delta\ge 0$.
The corresponding Gram matrix will be  $G=G(x,y,z)=W\,\Delta\,W^\top$.
Consider the
determinant $\delta$ of $\Delta$,  factorized in the following form
\begin{eqnarray}\nonumber
 \delta &\equiv &\det \Delta=-\frac{1}{4} (2 y+1) (2 x+2 y-1) (2 y+2 z-1)\\
 \label{K124a}
  &&\left(\left(y-\frac{1}{2}\right)^2-(x+1) (z+1)\right)
   \;.
\end{eqnarray}
It follows that ${\mathcal S}_{12}$ is bounded by the three planes defined by the vanishing of the first three linear factors
of $\delta$ and by the cone ${\mathcal C}$ defined by the vanishing of the last factor, see Figure \ref{FIGK2}.

The interior of ${\mathcal S}_{12}$ corresponds to $5$-dimensional ground states and its boundary mainly to $4$-dimensional ground states.
The exceptions that lead to physical ground state with dimensions two or three are indicated in Figure \ref{FIGK2}. They have been partially described
above in terms of independent rotations of subsets of spins. But there are other ``non-rotational" families: Three $1$-parameter families of $3$-dimensional
ground states joining the co-planar states are represented by the curves where the cone ${\mathcal C}$ is intersected by one of the three planes.
These conic sections turn out to be hyperbolas.
The existence of non-rotational families of ground states does not contradict the claim of \cite{CHS92} that the co-planar ground states
and the rotational families exhaust the set of all ground states of Kagome models with the boundary condition that all spins in surface triangles are co-planar.

We will give the family ${\mathbf s}(t),\;-1/2 \le t \le 1,$ joining $b_{12}$ and $c_{12}$ in closed form:
\begin{equation}\label{K125}
{\mathbf s}(t)=\left(
\begin{array}{ccc}
 1 & 0 & 0 \\
 -\frac{1}{2} & \frac{\sqrt{3}}{2} & 0 \\
 \frac{1}{t+1}-1 & -\frac{2 t+1}{\sqrt{3} (t+1)} &
   \frac{\sqrt{\frac{2}{3}} \sqrt{-2 t^2+t+1}}{t+1} \\
 -\frac{1}{2} & \frac{\sqrt{3}}{2} & 0 \\
 -\frac{1}{2} & -\frac{\sqrt{3}}{2} & 0 \\
 \frac{3}{2}-\frac{1}{t+1} & \frac{t-1}{2 \sqrt{3} (t+1)} &
   -\frac{\sqrt{\frac{2}{3}} \sqrt{-2 t^2+t+1}}{t+1} \\
 t & -\frac{t-1}{\sqrt{3}} & -\sqrt{\frac{2}{3}} \sqrt{-2 t^2+t+1} \\
 \frac{1}{2}-t & \frac{2 t+1}{2 \sqrt{3}} & \sqrt{\frac{2}{3}} \sqrt{-2
   t^2+t+1} \\
 t+\frac{1}{t+1}-2 & -\frac{t (t+2)}{\sqrt{3} (t+1)} &
   -\frac{\sqrt{\frac{2}{3}} t \sqrt{-2 t^2+t+1}}{t+1} \\
 \frac{1}{2}-t & \frac{2 t+1}{2 \sqrt{3}} & \sqrt{\frac{2}{3}} \sqrt{-2
   t^2+t+1} \\
 -\frac{1}{2} & -\frac{\sqrt{3}}{2} & 0 \\
 \frac{3}{2}-\frac{1}{t+1} & \frac{t-1}{2 \sqrt{3} (t+1)} &
   -\frac{\sqrt{\frac{2}{3}} \sqrt{-2 t^2+t+1}}{t+1} \\
\end{array}
\right).
\end{equation}

\begin{figure}[ht]
  \centering
    \includegraphics[width=1.0\linewidth]{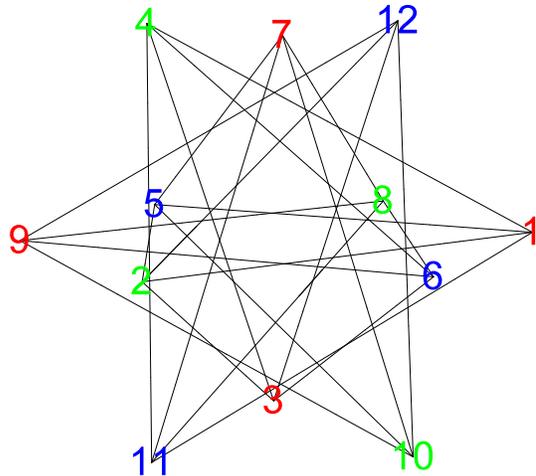}
  \caption
  {Visualisation of the $3$-dimensional ground state ${\mathbf e}$ as described in the text.
   }
  \label{FIGK3a}
\end{figure}

\begin{figure}[ht]
  \centering
    \includegraphics[width=1.0\linewidth]{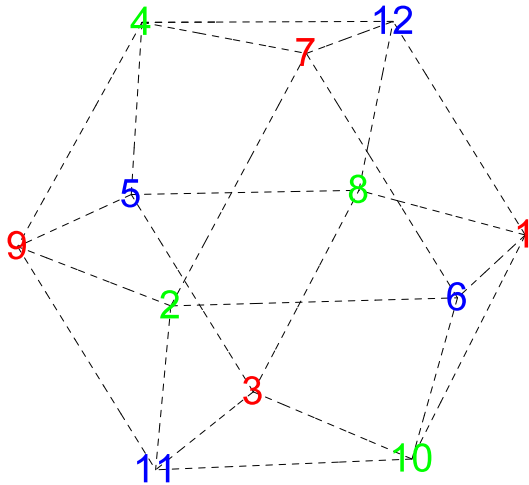}
  \caption
  {The same spin vectors as in Figure \ref{FIGK3a} but this time with the edges of the underlying cuboctadron. }
  \label{FIGK3b}
\end{figure}

Another $3$-dimensional ground state is represented by the vertex $e_{12}$ of the cone ${\mathcal C}$ that is isolated
within the set of physical ground states, see Figure \ref{FIGK2}. This ground state will be denoted
by ${\mathbf e}\equiv{\mathbf e}_{12}$ and can be written as
\begin{equation}\label{K126}
{\mathbf e} = \left(
\begin{array}{ccc}
 1 & 0 & 0 \\
 -\frac{1}{2} & -\frac{1}{\sqrt{2}} & \frac{1}{2} \\
 0 & 0 & -1 \\
 -\frac{1}{2} & \frac{1}{\sqrt{2}} & \frac{1}{2} \\
 -\frac{1}{2} & \frac{1}{\sqrt{2}} & -\frac{1}{2} \\
 \frac{1}{2} & -\frac{1}{\sqrt{2}} & \frac{1}{2} \\
 0 & 0 & 1 \\
 \frac{1}{2} & \frac{1}{\sqrt{2}} & -\frac{1}{2} \\
 -1 & 0 & 0 \\
 \frac{1}{2} & -\frac{1}{\sqrt{2}} & -\frac{1}{2} \\
 -\frac{1}{2} & -\frac{1}{\sqrt{2}} & -\frac{1}{2} \\
 \frac{1}{2} & \frac{1}{\sqrt{2}} & \frac{1}{2} \\
\end{array}
\right)
\end{equation}
It is interesting to represent the ground state ${\mathbf e}$ by drawing the numbers $\mu$ at the corresponding
position ${\mathbf e}_\mu$ and to join two numbers $\mu$ and $\nu$ if the corresponding spin sites are neighbors
in the spin system $K_{12}$, see Figure \ref{FIGK3a}.
We see that the spin vectors ${\mathbf e}_\mu,\;\mu=1,\ldots,12$ form the vertices of a cuboctahedron
if we now join two numbers in the case where ${\mathbf e}_\mu$ and ${\mathbf e}_\nu$ have a minimal positive distance, see Figure \ref{FIGK3b}.
(We hope that there will be no confusion between the ground state ${\mathbf e}={\mathbf e}_{12}$ and the vector ${\mathbf e}_\mu,\;\mu=12$.)
Together with the fact that $K_{12}$ is graph-theoretically isomorphic to the cuboctahedron this means that the ground state
${\mathbf e}$ can be understood as a permutation of the set of vertex vectors of the cuboctahedron. This has already been noted in \cite{SL03}
where the ground state ${\mathbf e}$ of the cuboctahedron has been discovered by group-theoretical methods.

In this context it is in order to mention that isolated three-dimensional ground states have also be recently obtained  in a slightly different setting, see  \cite{BZM17}. These authors have found numerical ground states of this kind in similar finite models of the Kagome lattice with periodic boundary conditions and  deviations of size $\delta$ from the uniform coupling and argue that these ground states should survive the limit $\delta\rightarrow 0$. In particular, according to \cite{B17} the above ground state ${\mathbf e}_{12}$ of $K_{12}$ has also been found numerically (up to an irrelevant rotation).
See also \cite{KRL17} for similar results.

From Figure \ref{FIGK3b} we may read that the part ${\sf Gr}$ of the symmetry group of $K_{12}$ that is isomorphic to the octahedral group
${\mathcal O}$ of order $24$ is generated by the two cyclic permutations $(4, 7, 12)(5, 2, 1)(9, 6, 8)(11, 10, 3)$ and
$(7, 6, 1, 12) (2, 10, 8, 4) (11, 3, 5, 9)$
corresponding to rotations of a triangle and a square, resp., within the cuboctahedron. The larger symmetry group ${\sf Gr_h}$
isomorphic to the octahedral group with reflections ${\mathcal O}_h$ of order $48$ is generated by additionally considering the permutation
$(1, 9)(2, 8)(3, 7)(4, 10)(5, 6)(11, 12)$ that interchanges each vertex with its antipode.

There is another Abelian group ${\sf T}$ of translations of $K_{12}$ that is generated by the two matrices $T_1$ and $T_2$ representing
even permutations $\pi_1=(1, 3)(2, 4)(5, 6)(7, 9)(8, 10)(11, 12)$ and $\pi_2=(1, 7), (2, 8), (3, 9), (4, 10), (5, 11), (6, 12)$.
It turns out that ${\sf T}$ is a subgroup of ${\sf Gr}$, as it must be since ${\mathcal O}_h$ is the maximal symmetry group of the cuboctahedron.
Recall that a Gram matrix is called ``${\sf G}$-symmetric" iff it commutes with a subgroup ${\sf G}$ of symmetries of the spin system. In our case
of $K_{12}$ we have to distinguish between ${\sf T}$-symmetry and ${\sf Gr_h}$-symmetry. It turns out that all Gram matrices $G(x,y,z)$ where
$(x,y,z)\in {\mathcal S}_{ADE}$ are ${\sf T}$-symmetric. On the other hand, the general ${\sf Gr_h}$-symmetry holds only for Gram matrices
of the form $G=G(x, -x/2,x),\;-1\le x\le 1$. This corresponds to the dashed black line joining the points $a_{12}$ and $e_{12}$ in Figure \ref{FIGK2}.
It follows that the eigenspaces of the corresponding Gram matrices, denoted by $G(\mathbf a)$ and $G({\mathbf e})$, carry irreducible representations of
${\mathcal O}$ customarily denoted by $E$ and $F_2$ of dimension two and three that together span the $5$-dimensional eigenspace of ${\mathbbm J}$
corresponding to the eigenvalue $j_{min}=-2$, see \cite{SL03}.

It will be illustrative to analyze the ${\sf T}$-symmetry of ground states in detail. As an example, we choose the state ${\mathbf e}={\mathbf e}_{12}$,
see (\ref{K126}),
and first consider $T_1\in{\sf T}$.
From $\left(T_1\,{\mathbf e}\right)\left({\mathbf e}^\top\,T_1^\top\right)=T_1\,G({\mathbf e})\,T_1^\top=G({\mathbf e})$
it follows that ${\mathbf e}$ and  $T_1\,{\mathbf e}$ have the same Gram matrix and hence, by Prop.~$4$ of \cite{S17}, they are $O(3)$-equivalent.
This means that there exists an $R_1\in O(3)$ such that $T_1\,{\mathbf e}={\mathbf e}\,R_1$. Indeed,
\begin{equation}\label{K127}
R_1=\left(
\begin{array}{ccc}
 0 & 0 & -1 \\
 0 & -1 & 0 \\
 -1 & 0 & 0 \\
\end{array}
\right)
\end{equation}
is the unique rotation matrix satisfying this requirement. $R_1$ is a rotation about the axis $(-1,0,1)$ with the angle $\pi$.
Thus one obtains the result that the ground state ${\mathbf e}$ has two periodic components, one parallel to $(-1,0,1)$ with the wave-number $q=0$,
and one perpendicular to $(-1,0,1)$ with the wave-number $q=\pi$.

An analogous result holds w.~r.~t.~the translation $T_2\in{\sf T}$ and the corresponding rotation matrix
\begin{equation}\label{K128}
R_2=\left(
\begin{array}{ccc}
 0 & 0 & 1 \\
 0 & -1 & 0 \\
 1 & 0 & 0 \\
\end{array}
\right)
\;.
\end{equation}
This time the $q=0$ component of ${\mathbf e}$ is parallel to $(1,0,1)$ and the $q=\pi$ component perpendicular to $(1,0,1)$.

\subsection{The $N=15$ Kagome model}\label{sec:K15}
\begin{figure}[ht]
  \centering
    \includegraphics[width=1.0\linewidth]{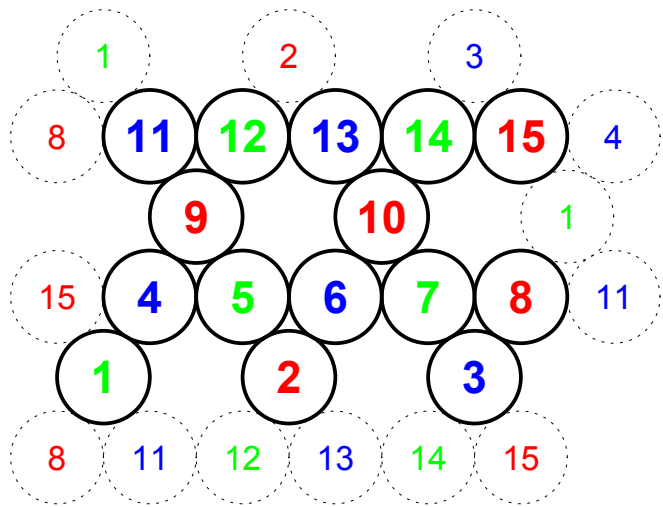}
  \caption
  {Representation of a finite $N=15$ model of the AF Kagome lattice. A co-planar ground state ${\mathbf f}_{15}$ with three spin directions forming mutual angles of
  $120^\circ$ is indicated by a coloring of the spin sites with the colors red, green, blue.
  }
  \label{FIGK15}
\end{figure}

The $N=15$ Kagome model $K_{15}$ is displayed in Figure \ref{FIGK15} together with a co-planar ground state ${\mathbf f}_{15}$. The
periodic extension ${\mathbf f}$ of ${\mathbf f}_{15}$ to the Kagome lattice has been called the ``Kosterlitz-Thouless-phase" , see, e.~g., \cite{CM13}.
It can be used to construct families of $3$-dimensional ground states in the following way: The double hexagon of spin sites with the numbers
$1,4,5,6,7,3,14,13,12,11$ and the central red spin $2$ is separated from the rest of the lattice by a collar of red spin sites.
Hence the spins within different double hexagons can be independently rotated about the red spin axes. Of course, the chosen unit cell
with $N=15$ spins is too small to represent these families of $3$-dimensional ground states.

The ${\mathbbm J}$-matrix of the Kagome model $K_{15}$  has a $6$-dimensional eigenspace corresponding to the lowest eigenvalue $j_{min}=-2$.
It is spanned by the $6$ columns of the matrix
\begin{equation}\label{K151}
W=\left(
\begin{array}{cccccc}
 0 & 0 & 0 & 0 & -1 & -1 \\
 0 & 0 & -1 & -1 & 0 & 0 \\
 -1 & -1 & 0 & 0 & 0 & 0 \\
 -1 & 0 & 0 & 0 & 1 & 1 \\
 1 & 0 & 0 & 1 & 0 & -1 \\
 -1 & 0 & 1 & 0 & 0 & 1 \\
 1 & 1 & 0 & 0 & 0 & -1 \\
 0 & 0 & 0 & 0 & 0 & 1 \\
 0 & 0 & 0 & -1 & -1 & 0 \\
 0 & -1 & -1 & 0 & 0 & 0 \\
 0 & 0 & 0 & 0 & 1 & 0 \\
 0 & 0 & 0 & 1 & 0 & 0 \\
 0 & 0 & 1 & 0 & 0 & 0 \\
 0 & 1 & 0 & 0 & 0 & 0 \\
 1 & 0 & 0 & 0 & 0 & 0 \\
\end{array}
\right).
\end{equation}
The solution set of the ADE (\ref{D4}) is also $6$-dimensional and difficult to analyze. Hence we consider the Abelian group ${\sf T}$ generated
by the linear representation $T_0$ of the cyclic permutation $(1,2,3,9,10)(4,6,8,12,14)(5,7,11,13,15)$
and confine ourselves to ${\sf T}$-symmetric ground states. It turns out
that the maximal symmetry group ${\sf Gr}$ of $K_{15}$ has the order $20$ and that each ${\sf T}$-symmetric ground state is also  ${\sf Gr}$-symmetric.
Moreover, ${\sf Gr}$ does not operate transitively on the spin sites but only on the subset with numbers $1, 2, 3, 9, 10$ and its complement.
 Nevertheless, it is correct to look for ground states that live in the eigenspace of
${\mathbbm J}$ corresponding to $j_{min}=-2$ and not to consider other gauges in the sense of \cite{S17}. The reason for this is that $j_{min}=-2$  is
equivalent to $E_{min}=-2\times 15 =-30$ as it must be for $10$ corner-sharing triangles in $K_{15}$ with a minimal energy of $-3$ for each triangle.

The convex set of ${\sf T}$-symmetric ground states can be parametrized by two parameters $u,v$ such that the general ${\sf T}$-symmetric solution
$\Delta$ of the ADE has the form
\begin{equation}\nonumber
\Delta=
\end{equation}
\begin{equation}\nonumber
\small
\left(
\begin{array}{cccccc}
 1 & -\frac{1}{2} & u & v & u-v-\frac{1}{2} & -u+v+1 \\
 -\frac{1}{2} & 1 & -\frac{1}{2} & u & v & u-v-\frac{1}{2} \\
 u & -\frac{1}{2} & 1 & -\frac{1}{2} & u & v \\
 v & u & -\frac{1}{2} & 1 & -\frac{1}{2} & u \\
 u-v-\frac{1}{2} & v & u & -\frac{1}{2} & 1 & -\frac{1}{2} \\
 -u+v+1 & u-v-\frac{1}{2} & v & u & -\frac{1}{2} & 1 \\
\end{array}
\right).
\end{equation}
\begin{equation}\label{K152}
 \end{equation}
The determinant $\delta(u,v)$ of $\Delta$ can be written in factorized form as
\begin{eqnarray}\nonumber
 \delta(u,v)&=&\frac{1}{256} (u-v) (3 u+v)\\
 \nonumber
 && \left(4 u+\left(2 \sqrt{5}-2\right) v+\sqrt{5}-5\right)^2\\
 \label{K153a}
 &&
   \left(-4 u+\left(2+2 \sqrt{5}\right) v+\sqrt{5}+5\right)^2.
\end{eqnarray}

\begin{figure}[ht]
  \centering
    \includegraphics[width=1.0\linewidth]{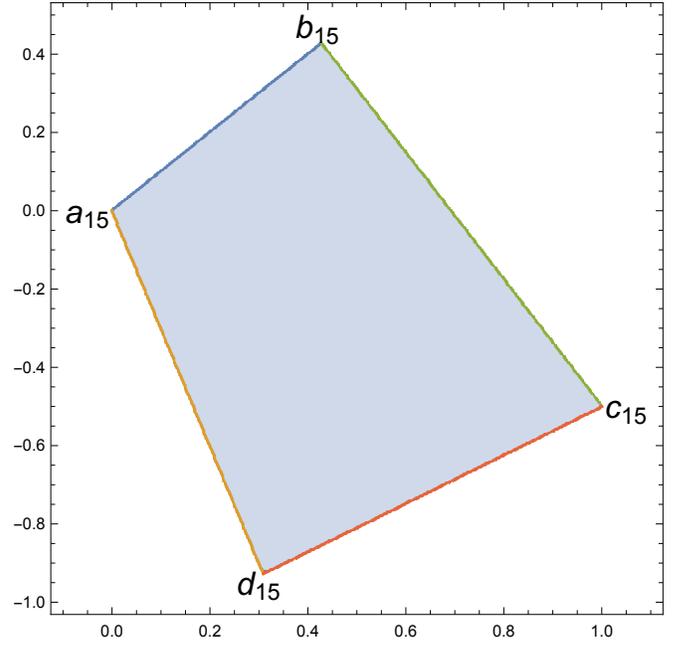}
  \caption
  {Representation of the convex set ${\mathcal S}_{ADE}^{sym}$ by a tetragon in the $u,v$-plane. The ground states corresponding to the vertices
  $a_{15},\,b_{15},\,c_{15},\,d_{15}$ are explained in the text.
  }
  \label{FIGK15A}
\end{figure}

Hence the convex set ${\mathcal S}_{ADE}^{sym}$ is isomorphic to the tetragon in the $u,v$-plane bounded by the $4$ lines defined by the vanishing of the
linear factors of $\delta(u,v)$, resp., see Figure \ref{FIGK15A}.
The vertices of the tetragon correspond to certain symmetric ground states of $K_{15}$. ${\mathbf a}_{15}$ is a $4$-dimensional state that will not be considered
further.  ${\mathbf c}_{15}$ generates the same the co-planar ground state of the Kagome lattice as the ground state ${\mathbf a}_{12}$ considered in
subsection \ref{sec:K12}. Note that the co-planar ground state ${\mathbf f}_{15}$ shown in Figure \ref{FIGK15} is not ${\sf T}$-symmetric.
The $3$-dimensional ground state ${\mathbf b}_{15}$ corresponds to the parameters $u=v=\frac{1}{4} \left(3 \sqrt{5}-5\right)\approx 0.427051$
and has the form
\vspace{3cm}

\begin{equation}\nonumber
 {\mathbf b}_{15}=
 \end{equation}
\begin{equation}\label{K154}
\left(
\begin{array}{ccc}
 -\frac{1}{2} & \frac{\sqrt{3}}{2} & 0 \\
 \frac{1}{4} \left(7-3 \sqrt{5}\right) & \frac{1}{4} \sqrt{3} \left(5 \sqrt{5}-9\right)
   & -\sqrt{\frac{3}{2} \left(13 \sqrt{5}-29\right)} \\
 \frac{1}{4} \left(7-3 \sqrt{5}\right) & \frac{1}{4} \sqrt{3} \left(\sqrt{5}-1\right) &
   -\sqrt{3 \left(\sqrt{5}-2\right)} \\
 1 & 0 & 0 \\
 -\frac{1}{2} & \sqrt{3}-\frac{\sqrt{15}}{2} & \sqrt{3 \left(\sqrt{5}-2\right)} \\
 \frac{1}{4} \left(3 \sqrt{5}-5\right) & \frac{1}{4} \sqrt{3} \left(5-3 \sqrt{5}\right)
   & -\sqrt{\frac{3}{2} \left(5 \sqrt{5}-11\right)} \\
 \frac{1}{4} \left(3 \sqrt{5}-5\right) & -\frac{1}{4} \sqrt{3} \left(\sqrt{5}-3\right)
   & \sqrt{3 \left(\sqrt{5}-2\right)} \\
 -\frac{1}{2} & -\frac{\sqrt{3}}{2} & 0 \\
 -\frac{1}{2} & \frac{1}{2} \sqrt{3} \left(\sqrt{5}-2\right) & -\sqrt{3
   \left(\sqrt{5}-2\right)} \\
 \frac{1}{2} \left(5-3 \sqrt{5}\right) & \sqrt{3} \left(\sqrt{5}-2\right) &
   -\sqrt{\frac{3}{2} \left(13 \sqrt{5}-29\right)} \\
 1 & 0 & 0 \\
 -\frac{1}{2} & \sqrt{3}-\frac{\sqrt{15}}{2} & \sqrt{3 \left(\sqrt{5}-2\right)} \\
 \frac{1}{4} \left(3 \sqrt{5}-5\right) & \frac{1}{4} \sqrt{3} \left(5-3 \sqrt{5}\right)
   & -\sqrt{\frac{3}{2} \left(5 \sqrt{5}-11\right)} \\
 \frac{1}{4} \left(3 \sqrt{5}-5\right) & -\frac{1}{4} \sqrt{3} \left(\sqrt{5}-3\right)
   & \sqrt{3 \left(\sqrt{5}-2\right)} \\
 -\frac{1}{2} & -\frac{\sqrt{3}}{2} & 0 \\
\end{array}
\right).
\end{equation}

In order to determine the wave numbers of ${\mathbf b}_{15}$ we calculate the unique rotation matrix $R\in O(3)$ satisfying
$T_0\,{\mathbf b}_{15}={\mathbf b}_{15}\,R$ as
\vspace{3cm}

\begin{equation}\nonumber
R=
\end{equation}
\begin{equation}\label{K1511}
\left(
\begin{array}{ccc}
 \frac{1}{4} \left(3 \sqrt{5}-5\right) & -\frac{1}{4} \sqrt{3} \left(3
   \sqrt{5}-5\right) & -\sqrt{\frac{3}{2} \left(5 \sqrt{5}-11\right)} \\
 -\frac{1}{4} \sqrt{3} \left(\sqrt{5}-3\right) & \frac{1}{4} \left(7 \sqrt{5}-13\right)
   & -\sqrt{\frac{5}{2} \left(5 \sqrt{5}-11\right)} \\
 \sqrt{3 \left(\sqrt{5}-2\right)} & \frac{1}{\sqrt{38+17 \sqrt{5}}} & 5-2 \sqrt{5} \\
\end{array}
\right).
\end{equation}

$R$ represents a rotation about the axis
$\left(\sqrt{\frac{1}{6} \left(1+\sqrt{5}\right)},-\sqrt{\frac{1}{2}
   \left(1+\sqrt{5}\right)},1\right)$
with an angle of $q=-\frac{2\pi}{5}$.

Analogously, ${\mathbf d}_{15}$ is the $3$-dimensional ground state corresponding to the parameters $u=\frac{1}{4} \left(\sqrt{5}-1\right)\approx 0.309017$
and $v=-\frac{3}{4} \left(\sqrt{5}-1\right)\approx -0.927051$
and has the form
\begin{equation}\nonumber
 {\mathbf d}_{15}=
 \end{equation}
\begin{equation}\label{K155}
\left(
\begin{array}{ccc}
 -\frac{1}{2} & \frac{\sqrt{3}}{2} & 0 \\
 \frac{1}{4} \left(3-\sqrt{5}\right) & \frac{1-3 \sqrt{5}}{4 \sqrt{3}} &
   -\sqrt{\frac{1}{6} \left(3 \sqrt{5}-5\right)} \\
 \frac{1}{4} \left(3-\sqrt{5}\right) & \frac{1+\sqrt{5}}{4 \sqrt{3}} &
   \frac{\sqrt[4]{5}}{\sqrt{3}} \\
 1 & 0 & 0 \\
 -\frac{1}{2} & -\frac{\sqrt{5}-2}{2 \sqrt{3}} & \frac{\sqrt[4]{5}}{\sqrt{3}} \\
 \frac{1}{4} \left(\sqrt{5}-1\right) & \frac{5 \left(\sqrt{5}-1\right)}{4 \sqrt{3}} &
   -\sqrt{\frac{1}{6} \left(7 \sqrt{5}-15\right)} \\
 -\frac{3}{4} \left(\sqrt{5}-1\right) & \frac{9-5 \sqrt{5}}{4 \sqrt{3}} & -\sqrt{3
   \sqrt{5}-\frac{20}{3}} \\
 \sqrt{5}-\frac{3}{2} & \sqrt{\frac{5}{3}}-\frac{5}{2 \sqrt{3}} & -\sqrt{\frac{14
   \sqrt{5}}{3}-10} \\
 -\frac{1}{2} & \frac{\sqrt{5}-2}{2 \sqrt{3}} & -\frac{\sqrt[4]{5}}{\sqrt{3}} \\
 \frac{1}{2} \left(\sqrt{5}-1\right) & -\frac{1}{\sqrt{3}} & \sqrt{\frac{1}{6} \left(3
   \sqrt{5}-5\right)} \\
 2-\sqrt{5} & -\frac{\sqrt{5}-1}{\sqrt{3}} & \sqrt{\frac{14 \sqrt{5}}{3}-10} \\
 \sqrt{5}-\frac{3}{2} & \frac{\sqrt{\frac{5}{3}}}{2} & \sqrt{3 \sqrt{5}-\frac{20}{3}}
   \\
 -\frac{3}{4} \left(\sqrt{5}-1\right) & \frac{\sqrt{5}-1}{4 \sqrt{3}} & \sqrt{\frac{7
   \sqrt{5}}{6}-\frac{5}{2}} \\
 \frac{1}{4} \left(\sqrt{5}-1\right) & -\frac{\sqrt{5}-5}{4 \sqrt{3}} &
   -\frac{\sqrt[4]{5}}{\sqrt{3}} \\
 -\frac{1}{2} & -\frac{\sqrt{3}}{2} & 0 \\
\end{array}
\right).
\end{equation}

In order to determine the wave numbers of ${\mathbf d}_{15}$ we calculate the unique rotation matrix $R\in O(3)$ satisfying
$T_0\,{\mathbf d}_{15}={\mathbf d}_{15}\,R$ as

\begin{equation}\nonumber
R=
\end{equation}
\begin{equation}\label{K1512}
\left(
\begin{array}{ccc}
 \frac{1}{4} \left(\sqrt{5}-1\right) & \frac{5 \left(\sqrt{5}-1\right)}{4 \sqrt{3}} &
   -\sqrt{\frac{1}{6} \left(7 \sqrt{5}-15\right)} \\
 -\frac{\sqrt{5}-5}{4 \sqrt{3}} & \frac{1}{12} \left(-3-\sqrt{5}\right) & -\frac{1}{3}
   \sqrt{\frac{1}{2} \left(5+3 \sqrt{5}\right)} \\
 -\frac{\sqrt[4]{5}}{\sqrt{3}} & \frac{1}{3} \sqrt[4]{5} \left(\sqrt{5}-2\right) &
   \frac{1}{3} \left(3-2 \sqrt{5}\right) \\
\end{array}
\right).
\end{equation}

$R$ represents a rotation about the axis
$\left(-\sqrt{\frac{3}{10} \left(5+3 \sqrt{5}\right)},\frac{2
   \sqrt[4]{5}}{\sqrt{5}-5},1\right)$
with an angle of $q=-\frac{4\pi}{5}$.

\subsection{The $N=18$ Kagome model}\label{sec:K18}
\begin{figure}[ht]
  \centering
    \includegraphics[width=1.0\linewidth]{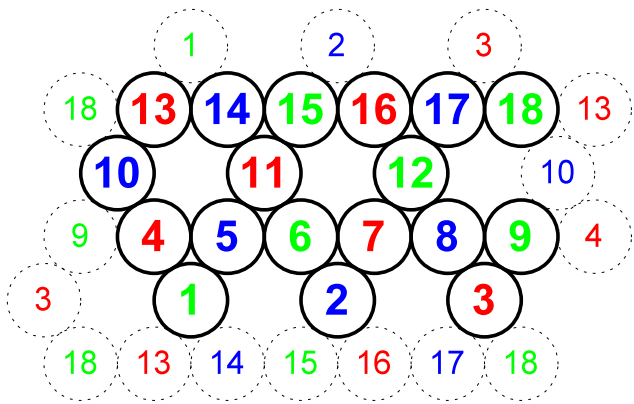}
  \caption
  { {Representation of a finite $N=18$ model of the AF Kagome lattice. A co-planar ground state ${\mathbf a}_{18}$
  with three spin directions forming mutual angles of
  $120^\circ$ is indicated by a coloring of the spin sites with the colors red, green, blue.
  } }
  \label{FIGK4}
\end{figure}

In the literature on the Kagome lattice another class of ground states has been discussed that is generated by the so-called
$ \sqrt{3}\times\sqrt{3}$-structure living in a unit cell of $9$ spins, see, e.~g., \cite{CM13}. In order to account for this class of ground states
we consider another Kagome model $K_{18}$ and a co-planar ground state denoted by ${\mathbf a}_{18}$ indicated by a coloring of the spin sites, see
Figure \ref{FIGK4}.\\

Again, it is possible to perform independent rotations of subsets of spin vectors. As one can see in the Figure \ref{FIGK4},
the hexagon $(6,7,12,16,15,11)$ is separated from the hexagon $(1,4,9,3,18,13)$ by a set of blue spins. Hence one can independently rotate the
spin vectors of both hexagons about the blue spin axis. This yields a $1$-parameter family of $3$-dimensional ground states.
After a rotation of $180^\circ$ a new co-planar ground state ${\mathbf b}_{18}$ is generated. Similarly, it
is possible to construct two more families of ground states by independent rotations within hexagons surrounded by green or red spins
and to obtain new co-planar ground states ${\mathbf c}_{18}$ and ${\mathbf d}_{18}$ as a by-product.

The application of the theory outlined in \cite{S17} is more difficult than in the case of $K_{12}$ due to the larger $N$. The convex set
${\mathcal S}_{18}$ will depend on ten real parameters and is not easy to analyze. Hence we have decided to confine ourselves to the subset
of ${\sf T}$-symmetric ground states. Here ${\sf T}$ is the Abelian group of translations generated by the linear representation $T_1$ of
the cyclic permutation $\tau=(1, 2, 3)(4, 6, 8)(5, 7, 9)(10, 11, 12)(13, 15, 17)(14, 16, 18)$. Note that the translation into the perpendicular
direction is represented by the unit matrix due to our choice of the unit cell in $K_{18}$. It is plausible and has been directly confirmed
that the ground states considered in the last paragraphs are not ${\sf T}$-symmetric with the exception of the co-planar ground state ${\mathbf a}_{12}$.
Hence if we find $3$-dimensional ${\sf T}$-symmetric ground states of $K_{18}$ these will be clearly different from the ground states considered above.

There  exists an even larger Abelian group of symmetries generated by the cyclic permutation
$(1,11,2,12,3,10)(4,14,6,16,8,18)(5,15,7,17,9,13)$,
but the among the $3$-dimensional ground states that are symmetric w.~r.~t.~this larger group there are less new ones. Hence we prefer to work with the above-defined group ${\sf T}$.

The ${\mathbbm J}$-matrix of the Kagome model $K_{18}$  has a $7$-dimensional eigenspace corresponding to the lowest eigenvalue $j_{min}=-2$.
It is spanned by the $7$ columns of the matrix
\begin{equation}\label{K180}
W=\left(
\begin{array}{ccccccc}
 0 & 0 & 0 & 0 & -1 & -1 & 0 \\
 0 & 0 & -1 & -1 & 0 & 0 & 0 \\
 -1 & -1 & 0 & 0 & 0 & 0 & 0 \\
 1 & 0 & 0 & 0 & 0 & 1 & -1 \\
 -1 & 0 & 0 & 0 & 1 & 0 & 1 \\
 1 & 0 & 0 & 1 & 0 & 0 & -1 \\
 -1 & 0 & 1 & 0 & 0 & 0 & 1 \\
 1 & 1 & 0 & 0 & 0 & 0 & -1 \\
 0 & 0 & 0 & 0 & 0 & 0 & 1 \\
 -1 & 0 & 0 & 0 & 0 & -1 & 0 \\
 0 & 0 & 0 & -1 & -1 & 0 & 0 \\
 0 & -1 & -1 & 0 & 0 & 0 & 0 \\
 0 & 0 & 0 & 0 & 0 & 1 & 0 \\
 0 & 0 & 0 & 0 & 1 & 0 & 0 \\
 0 & 0 & 0 & 1 & 0 & 0 & 0 \\
 0 & 0 & 1 & 0 & 0 & 0 & 0 \\
 0 & 1 & 0 & 0 & 0 & 0 & 0 \\
 1 & 0 & 0 & 0 & 0 & 0 & 0 \\
\end{array}
\right).
\end{equation}

As mentioned above, the solutions of the ADE depend on ten real parameters. The subset of ${\sf T}$-symmetric solutions $\Delta(u,x,y,z)$
is still characterized by a convex set ${\mathcal S}^{sym}_{18}$ of dimension $4$ and thus cannot be represented graphically.
The general ${\sf T}$-symmetric solution of the ADE assumes the form
\begin{equation}\nonumber
\Delta(u,x,y,z)=
\end{equation}
\begin{eqnarray}\nonumber
\begin{array}{ccccccc}
 1 & -\frac{1}{2} & u & x & u & -\frac{1}{2} & y \\
 -\frac{1}{2} & 1 & -\frac{1}{2} & z & x & z & \frac{1}{2}-y \\
 u & -\frac{1}{2} & 1 & -\frac{1}{2} & u & x & u+y-1 \\
 x & z & -\frac{1}{2} & 1 & -\frac{1}{2} & z & x-y+1 \\
 u & x & u & -\frac{1}{2} & 1 & -\frac{1}{2} & u+y-1 \\
 -\frac{1}{2} & z & x & z & -\frac{1}{2} & 1 & \frac{1}{2}-y \\
 y & \frac{1}{2}-y & u+y-1 & x-y+1 & u+y-1 & \frac{1}{2}-y & 1 \\
\end{array}
&&\\
&&\label{K181}
\end{eqnarray}
Its determinant $\delta(u,x,y,z)$ can be written in factorized form as
\begin{eqnarray}\nonumber
\delta(u,x,y,z)&=&-\frac{1}{8} (y-1) \left(4 u z-4 u-4 x^2-4 x-4 z+3\right)^2 \\
\nonumber
&&\left(3 u y+4 u    z-u-x^2+3 x y-x+3 y z-z\right)\\
\label{K182}
&\equiv& -\frac{1}{8} (y-1) \,\delta_2^2\,\delta_3\;.
\end{eqnarray}
Hence ${\mathcal S}^{sym}_{18}$ is bounded by a $3$-dimensional face ${\mathcal F}_{18}$ defined by $y=1$ and two $3$-dimensional hyper-surfaces
${\mathcal H}_2$ and ${\mathcal H}_3$ defined by the vanishing of $\delta_2$ and $\delta_3$, resp..
(More precisely, the vanishing of $\delta_2$ and $\delta_3$ defines real, algebraic varieties that are extended beyond the boundary of
${\mathcal S}^{sym}_{18}$. In the following we implicitly understand by ${\mathcal H}_2$ and ${\mathcal H}_3$  the restriction of these
varieties to the boundary of ${\mathcal S}^{sym}_{18}$.) The points in the
interior of ${\mathcal S}^{sym}_{18}$ correspond to $7$-dimensional ground states. Since we are mainly interested in physical
ground states of dimension two or three it is in order to more closely investigate the boundary of ${\mathcal S}^{sym}_{18}$.
It is very plausible and has been checked by examples that the rank of Gram matrices (dimension of ground states)
corresponding to the points of ${\mathcal S}^{sym}_{18}$ decreases by $1$ if one reaches the boundary of ${\mathcal S}^{sym}_{18}$
at some interior point of ${\mathcal F}_{18}$ or ${\mathcal H}_3$, and by $2$ at interior points of ${\mathcal H}_2$. The latter is due
to the factor $\delta_2^2$ in the determinant (\ref{K182}). This implies that Gram matrices of rank $3$ should occur at the intersection
${\mathcal F}_{18} \cap {\mathcal H}_2 \cap {\mathcal H}_3$ and this is indeed the case, see below. Another possibility for the occurrence
of Gram matrices of rank $3$ are singular points of ${\mathcal H}_2$, where the rank loss of $G$ may assume the value $4$, see below.
We start with an investigation of  the face ${\mathcal F}_{18}$.

\subsubsection{The face ${\mathcal F}_{18}$}\label{sec:K181}
We insert $y=1$ into $\delta_2$ and $\delta_3$ and obtain
\begin{eqnarray}\label{K183a}
  \delta_2 &=& 4\left( (z-1)(u-1)-(x+1/2)^2 \right)\;, \\
  \label{K183b}
  \delta_3 &=& 4(z+1/2)(u+1/2)-(x-1)^2\;.
\end{eqnarray}

\begin{figure}[ht]
  \centering
    \includegraphics[width=1.0\linewidth]{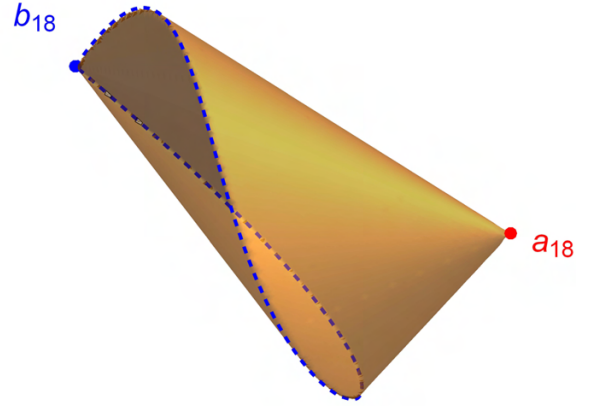}
  \caption
  {Representation of the face ${\mathcal F}_{18}$ of ${\mathcal S}^{sym}_{18}$
  bounded by two elliptical cones ${\mathcal C}_2$ and ${\mathcal C}_3$.
  The intersection (dashed blue curve) yields a $1$-parameter family ${\mathcal L}_1$ of $3$-dimensional ground states ${\boldsymbol \ell}(t)$ joining the
  co-planar ground state ${\mathbf b}_{18}$ with itself. The vertex $a_{18}$ of the cone ${\mathcal C}_2$  (red point) corresponds
  to a co-planar ground state ${\mathbf a}_{18}$ that is periodically extended to the same state on the Kagome lattice as the co-planar ground state
  ${\mathbf a}_{12}$ of $K_{12}$, see Figure \ref{FIGK1}.
  }
  \label{FIGK5}
\end{figure}

Hence the boundary of ${\mathcal F}_{18}$ is formed by two elliptical cones ${\mathcal C}_2$ and ${\mathcal C}_3$ defined
by the vanishing of $\delta_2$ and $\delta_3$, resp., such that the vertex of ${\mathcal C}_3$ lies on ${\mathcal C}_2$, see
Figure \ref{FIGK5}.

The intersection of the two cones ${\mathcal C}_2$ and ${\mathcal C}_3$ yields a $1$-parameter curve ${\mathcal L}_1$ of $3$-dimensional ground states
joining the co-planar ground state ${\mathbf b}_{18}$ with itself, see Figure \ref{FIGK5}. ${\mathbf b}_{18}$ corresponds simultaneously to the vertex of the
cone ${\mathcal C}_3$. Interestingly, the vertex $a_{18}$ of the other cone ${\mathcal C}_2$ corresponds to a co-planar ground state ${\mathbf a}_{18}$ that we have already encountered: If periodically extended to the Kagome lattice it coincides with the co-planar ground state ${\mathbf a}_{12}$ of the Kagome model
$K_{12}$, see subsection \ref{sec:K12}.

The $1$-parameter curve ${\mathcal L}_1$ can be described by the parametrization
\begin{eqnarray}\label{K184a}
u(t)&=&\frac{-6\, t^2+4 \sqrt{3}\, t-1}{2 \left(2-\sqrt{3}\, t\right)^2},\\
\label{K184b}
x(t)&=&\frac{-3\, t^2+\sqrt{3}\, t+1}{\sqrt{3}\, t-2},\\
\label{K184c}
z(t)&=&1-\frac{3\, t^2}{2},
\end{eqnarray}
where $-1\le t\le 1$. The corresponding ground states ${\boldsymbol \ell}(t)$ can be calculated in closed form.
We will shortly explain how. Inserting $y=1$ and the parametrization (\ref{K184a})--(\ref{K184c}) into (\ref{K181}) we
obtain $G(t)=W\, \Delta(u(t),x(t),1,z(t))\,W^\top$, where $W$ is given by (\ref{K180}). The explicit result
for $G(t)$ is too complex to be represented here. To obtain ${\boldsymbol \ell}(t)$
we choose three spin vectors of the ground state ${\boldsymbol \ell}(t)$ according to
\begin{eqnarray}\label{K185a}
  {\boldsymbol \ell}_5 &=& \left(1,0,0\right), \\
  \label{K185b}
 {\boldsymbol \ell}_1 &=& \left( -\frac{1}{2}, \frac{\sqrt{3}}{2} , 0\right),\\
  \label{K185c}
  {\boldsymbol \ell}_6(t) &=&\left( -\frac{1}{2} ,\frac{1}{2} \sqrt{3} \left(2 t^2-1\right) , t \sqrt{3-3 t^2}\right).
\end{eqnarray}
This choice is compatible with
$G_{1,5}(t)=G_{5,6}(t)= \frac{1}{2}$ and $G_{1,6}(t)= \frac{1}{2} \left(3 t^2-1\right)$.
The $3$ vectors (\ref{K185a}) -- (\ref{K185c}) form a basis in ${\mathbbm R}^3$ for $-1<t<1$.
The scalar products of an arbitrary vector ${\mathbf x}\in{\mathbbm R}^3$ with these $3$ basis vectors
can be written as a vector ${\boldsymbol \xi}\in{\mathbbm R}^3$
that is obtained from ${\mathbf x}$ by  ${\boldsymbol \xi}=Q\,{\mathbf x}$, where $Q$ denotes the matrix
\begin{equation}\label{K186}
 Q\equiv \left(
\begin{array}{c}
{\boldsymbol \ell}_5\\
{\boldsymbol \ell}_1\\
{\boldsymbol \ell}_6(t)
\end{array}
\right).
\end{equation}
Hence ${\mathbf x}=Q^{-1}\,{\boldsymbol \xi}$ with
\begin{equation}\label{K187}
Q^{-1}=\left(
\begin{array}{ccc}
 1 & 0 & 0 \\
 \frac{1}{\sqrt{3}} & \frac{2}{\sqrt{3}} & 0 \\
 \frac{\sqrt{1-t^2}}{\sqrt{3} t} & \frac{1-2 t^2}{t \sqrt{3-3 t^2}} &
   \frac{1}{t \sqrt{3-3 t^2}} \\
\end{array}
\right).
\end{equation}
We apply these equations to the vector ${\mathbf x}={\boldsymbol \ell}_\mu(t)$. The vector of scalar products with the $3$ basis vectors is
\begin{equation}\label{K188}
{\boldsymbol \xi}=\left(
\begin{array}{c}
G_{\mu 5}(t)\\
G_{\mu 1}(t)\\
G_{\mu 6}(t)
\end{array}
\right).
\end{equation}
Hence the cartesian components of ${\boldsymbol \ell}_\mu(t)$ are given by
\begin{equation}\label{K187}
 {\boldsymbol \ell}_\mu(t)=Q^{-1}\,
 \left(
\begin{array}{c}
G_{\mu 5}(t)\\
G_{\mu 1}(t)\\
G_{\mu 6}(t)
\end{array}
\right).
\end{equation}
These spin vectors satisfy ${\boldsymbol \ell}_\mu(t)\cdot{\boldsymbol \ell}_\nu(t)=G_{\mu\nu}(t)$ for $\mu,\nu=1,\ldots,18$.
Their explicit form is
\onecolumngrid
\begin{eqnarray}\nonumber
{\boldsymbol \ell}(t)&=&
\frac{1}{\left(2 \sqrt{3}-3 t\right)^2}\\
\nonumber
&&\left(
\begin{array}{ccc}
 -\frac{1}{2} \left(2 \sqrt{3}-3 t\right)^2 & \frac{1}{2} \sqrt{3} \left(2
   \sqrt{3}-3 t\right)^2 & 0 \\
 \frac{3}{2} \left(9 t^2-8 \sqrt{3} t+5\right) & \frac{3}{2} \sqrt{3} \left(-6
   t^2+4 \sqrt{3} t-1\right) \left(t^2-1\right) & 3 \sqrt{1-t^2} \left(t
   \left(-3 \sqrt{3} t^2+6 t+\sqrt{3}\right)-3\right) \\
 3 t \left(3 t \left(\sqrt{3} t-2\right)-\sqrt{3}\right)+\frac{15}{2} &
   \frac{3}{2} \left(\sqrt{3}-2 t \left(t \left(3 t \left(\sqrt{3}
   t-3\right)+\sqrt{3}\right)+3\right)\right) & 3 \sqrt{1-t^2} \left(-3
   \sqrt{3} t^3+9 t^2-4 \sqrt{3} t+3\right) \\
 -\frac{1}{2} \left(2 \sqrt{3}-3 t\right)^2 & -\frac{1}{2} \sqrt{3} \left(2
   \sqrt{3}-3 t\right)^2 & 0 \\
 \left(2 \sqrt{3}-3 t\right)^2 & 0 & 0 \\
 -\frac{1}{2} \left(2 \sqrt{3}-3 t\right)^2 & \frac{1}{2} \sqrt{3} \left(2
   \sqrt{3}-3 t\right)^2 \left(2 t^2-1\right) & \left(2 \sqrt{3}-3 t\right)^2 t
   \sqrt{3-3 t^2} \\
 -9 t^2+6 \sqrt{3} t-\frac{3}{2} & 3 \left(6 t-5 \sqrt{3}\right) t^2+\frac{9
   \sqrt{3}}{2} & 3 \sqrt{1-t^2} \left(6 t^2-5 \sqrt{3} t+3\right) \\
 -3 \left(t \left(3 t \left(\sqrt{3} t-3\right)+\sqrt{3}\right)+2\right) & 3
   \left(t^2-1\right) \left(3 t \left(\sqrt{3} t-3\right)+2 \sqrt{3}\right) & 3
   t \left(3 t-\sqrt{3}\right) \left(\sqrt{3} t-2\right) \sqrt{1-t^2} \\
 -9 t^2+6 \sqrt{3} t-\frac{3}{2} & 6 t \left(\sqrt{3} t-3\right)+\frac{9
   \sqrt{3}}{2} & 3 \left(2 \sqrt{3} t-3\right) \sqrt{1-t^2} \\
 \frac{3}{2} \left(9 t^2-8 \sqrt{3} t+5\right) & -\frac{3}{2} \sqrt{3}
   \left(t^2-1\right) & 3 \left(3-2 \sqrt{3} t\right) \sqrt{1-t^2} \\
 -\frac{1}{2} \left(2 \sqrt{3}-3 t\right)^2 & \frac{1}{2} \sqrt{3} \left(2
   \sqrt{3}-3 t\right)^2 \left(1-2 t^2\right) & -\left(2 \sqrt{3}-3 t\right)^2
   t \sqrt{3-3 t^2} \\
 3 t \left(3 t \left(\sqrt{3} t-2\right)-\sqrt{3}\right)+\frac{15}{2} &
   \frac{3}{2} \left(6 t \left(-\sqrt{3} t^3+t^2+2 \sqrt{3}
   t-3\right)+\sqrt{3}\right) & 9 \left(\sqrt{3} t-1\right)
   \left(1-t^2\right)^{3/2} \\
 -\frac{1}{2} \left(2 \sqrt{3}-3 t\right)^2 & -\frac{1}{2} \sqrt{3} \left(2
   \sqrt{3}-3 t\right)^2 & 0 \\
 \left(2 \sqrt{3}-3 t\right)^2 & 0 & 0 \\
 -\frac{1}{2} \left(2 \sqrt{3}-3 t\right)^2 & \frac{1}{2} \sqrt{3} \left(2
   \sqrt{3}-3 t\right)^2 \left(2 t^2-1\right) & \left(2 \sqrt{3}-3 t\right)^2 t
   \sqrt{3-3 t^2} \\
 -9 t^2+6 \sqrt{3} t-\frac{3}{2} & 3 \left(6 t-5 \sqrt{3}\right) t^2+\frac{9
   \sqrt{3}}{2} & 3 \sqrt{1-t^2} \left(6 t^2-5 \sqrt{3} t+3\right) \\
 -3 \left(t \left(3 t \left(\sqrt{3} t-3\right)+\sqrt{3}\right)+2\right) & 3
   \left(t^2-1\right) \left(3 t \left(\sqrt{3} t-3\right)+2 \sqrt{3}\right) & 3
   t \left(3 t-\sqrt{3}\right) \left(\sqrt{3} t-2\right) \sqrt{1-t^2} \\
 -9 t^2+6 \sqrt{3} t-\frac{3}{2} & 6 t \left(\sqrt{3} t-3\right)+\frac{9
   \sqrt{3}}{2} & 3 \left(2 \sqrt{3} t-3\right) \sqrt{1-t^2} \\
\end{array}
\right).\\
&&\label{K1810}
\end{eqnarray}

\twocolumngrid

In order to determine the wave numbers of ${\boldsymbol \ell}(t)$ we calculate the unique rotation matrix $R(t)\in O(3)$ satisfying
$T_1\,{\boldsymbol \ell}(t)={\boldsymbol \ell}(t)\,R(t)$ as
\onecolumngrid
\begin{equation}\label{K1811}
R(t)=\frac{1}{\left(2-\sqrt{3} t\right)^2}
\left(
\begin{array}{ccc}
 -3 t^2+2 \sqrt{3} t-\frac{1}{2} & \frac{1}{2} \sqrt{3} \left(\sqrt{3}-2
   t\right)^2 & \left(2 \sqrt{3} t-3\right) \sqrt{1-t^2} \\
 6 t^3-5 \sqrt{3} t^2+\frac{3 \sqrt{3}}{2} & -6 t^4+6 \sqrt{3} t^3-4 \sqrt{3}
   t+\frac{5}{2} & \sqrt{1-t^2} \left(\sqrt{3}-6 t \left(t^2-\sqrt{3}
   t+1\right)\right) \\
 \sqrt{1-t^2} \left(6 t^2-5 \sqrt{3} t+3\right) & -\sqrt{1-t^2} \left(3 t
   \left(2 t \left(t-\sqrt{3}\right)+1\right)+\sqrt{3}\right) & t \left(3 t
   \left(2 t \left(t-\sqrt{3}\right)+1\right)+2 \sqrt{3}\right)-2 \\
\end{array}
\right).
\end{equation}
\twocolumngrid

$R(t)$ represents a rotation about the axis
$\left(\sqrt{1-t^2},\left(\sqrt{3}-2 t\right) \sqrt{1-t^2},-\left(\sqrt{3}-2 t\right) t\right)$
with an angle of $q=\frac{2\pi}{3}$.

\subsubsection{The hyper-surface ${\mathcal H}_{2}$}\label{sec:K182}
As mentioned above, the hyper-surface ${\mathcal H}_{2}$ is the intersection of the set of solutions of
\begin{equation}\label{K1812}
 \delta_2\equiv \left(4 u z-4 u-4 x^2-4 x-4 z+3\right)=0
\end{equation}
with the boundary of ${\mathcal S}^{sym}_{18}$. We expect additional $3$-dimensional ground states corresponding to the singular points
of ${\mathcal H}_{2}$, that are characterized by the vanishing of the gradient $\nabla\,\delta_2$. It turns out the the solution of the
simultaneous equations $\delta_2=0$ and $\nabla\,\delta_2={\mathbf 0}$ is the $1$-dimensional family given by $u=1,\;x=-1/2,\;z=1$ and $y\in{\mathbbm R}$.
$\Delta(1,-1/2,y,-1/2)$ is positively semi-definite for $-1/2\le y\le 1$ and has the rank $3$ for $-1/2< y< 1$. The calculation of the
corresponding family of $3$-dimensional ground states ${\mathbf g}(y)$ is completely analogous to that in subsection \ref{sec:K181}.
Hence it suffices to give the final result:
\begin{equation}\label{K1813}
{\mathbf g}(y)=
\left(
\begin{array}{ccc}
 1 & 0 & 0 \\
 1 & 0 & 0 \\
 1 & 0 & 0 \\
 -\frac{1}{2} & \frac{\sqrt{3}}{2} & 0 \\
 -\frac{1}{2} & -\frac{\sqrt{3}}{2} & 0 \\
 -\frac{1}{2} & \frac{\sqrt{3}}{2} & 0 \\
 -\frac{1}{2} & -\frac{\sqrt{3}}{2} & 0 \\
 -\frac{1}{2} & \frac{\sqrt{3}}{2} & 0 \\
 -\frac{1}{2} & -\frac{\sqrt{3}}{2} & 0 \\
 1 & 0 & 0 \\
 1 & 0 & 0 \\
 1 & 0 & 0 \\
 -\frac{1}{2} & \frac{4 y-1}{2 \sqrt{3}} & \sqrt{\frac{2}{3}} \sqrt{(1-y) (2
   y+1)} \\
 -\frac{1}{2} & -\frac{4 y-1}{2 \sqrt{3}} & -\sqrt{\frac{2}{3}} \sqrt{(1-y)
   (2 y+1)} \\
 -\frac{1}{2} & \frac{4 y-1}{2 \sqrt{3}} & \sqrt{\frac{2}{3}} \sqrt{(1-y) (2
   y+1)} \\
 -\frac{1}{2} & -\frac{4 y-1}{2 \sqrt{3}} & -\sqrt{\frac{2}{3}} \sqrt{(1-y)
   (2 y+1)} \\
 -\frac{1}{2} & \frac{4 y-1}{2 \sqrt{3}} & \sqrt{\frac{2}{3}} \sqrt{(1-y) (2
   y+1)} \\
 -\frac{1}{2} & -\frac{4 y-1}{2 \sqrt{3}} & -\sqrt{\frac{2}{3}} \sqrt{(1-y)
   (2 y+1)} \\
\end{array}
\right).
\end{equation}
It follows that this family of ground states is essentially the same as that described at the beginning of section \ref{sec:K12}
with a wave number $q=0$ resulting from an independent rotation of every second row of spins. This is somewhat disappointing
at first sight since we were looking for new ground states.  On the other hand this finding confirms that the present method is
suited to find all (symmetric) ground states even if they appear trivial.

\subsubsection{The hyper-surface ${\mathcal H}_{3}$}\label{sec:K183}

\begin{figure}[ht]
  \centering
    \includegraphics[width=1.0\linewidth]{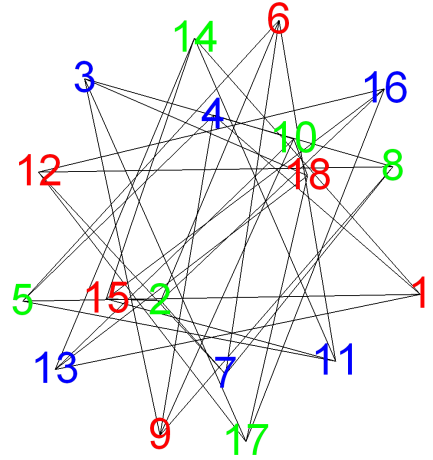}
  \caption
  {Representation of the $3$-dimensional ground state ${\mathbf e}_{18}$ defined in (\ref{K1815}).
  }
  \label{FIGK6}
\end{figure}

As mentioned above, the hyper-surface ${\mathcal H}_{3}$ is the intersection of the set of solutions of
\begin{equation}\label{K1814}
\delta_3\equiv \left(3 u y+4 u    z-u-x^2+3 x y-x+3 y z-z\right)=0
\end{equation}
with the boundary of ${\mathcal S}^{sym}_{18}$. We are looking for additional $3$-dimensional ground states corresponding to the singular points
of ${\mathcal H}_{3}$, that are characterized by the vanishing of the gradient $\nabla\,\delta_3$. It turns out the the solution of the
simultaneous equations $\delta_3=0$ and $\nabla\,\delta_3={\mathbf 0}$ is the $1$-dimensional family given by
$u=1/4 (1 - 3 y),\;x= 1/2 (-1 + 3 y),\;z=1/4 (1 - 3 y)$ and $y\in{\mathbbm R}$.
$\Delta(1/4 (1 - 3 y),1/2 (-1 + 3 y),y,1/4 (1 - 3 y))$ is positively semi-definite for $-1/3\le y\le 1$ and has the rank $5$ for $-1/3< y< 1$.
For the end-points of this family the rank of the Gram matrices decreases to $2$ at $y=1$ and to $3$ at $y=-1/3$.
The co-planar ground state corresponding to $y=1$ is again the ground state ${\mathbf a}_{18}$ displayed in Figure \ref{FIGK4}.
Hence we will concentrate of the $3$-dimensional ground state ${\mathbf e}_{18}$ corresponding to $y=-1/3$. It can be calculated
explicitly:
\begin{equation}\label{K1815}
{\mathbf e}_{18}=
\left(
\begin{array}{ccc}
 1 & 0 & 0 \\
 -\frac{1}{2} & \frac{1}{2 \sqrt{3}} & -\sqrt{\frac{2}{3}} \\
 -\frac{1}{2} & -\frac{1}{2 \sqrt{3}} & \sqrt{\frac{2}{3}} \\
 -\frac{1}{2} & \frac{\sqrt{3}}{2} & 0 \\
 -\frac{1}{2} & -\frac{\sqrt{3}}{2} & 0 \\
 0 & \frac{1}{\sqrt{3}} & \sqrt{\frac{2}{3}} \\
 \frac{1}{2} & -\frac{\sqrt{3}}{2} & 0 \\
 \frac{1}{2} & \frac{\sqrt{3}}{2} & 0 \\
 0 & -\frac{1}{\sqrt{3}} & -\sqrt{\frac{2}{3}} \\
 \frac{1}{2} & -\frac{1}{2 \sqrt{3}} & \sqrt{\frac{2}{3}} \\
 \frac{1}{2} & \frac{1}{2 \sqrt{3}} & -\sqrt{\frac{2}{3}} \\
 -1 & 0 & 0 \\
 -\frac{1}{2} & -\frac{7}{6 \sqrt{3}} & -\frac{2
   \sqrt{\frac{2}{3}}}{3} \\
 -\frac{1}{2} & \frac{7}{6 \sqrt{3}} & \frac{2 \sqrt{\frac{2}{3}}}{3}
   \\
 0 & -\frac{5}{3 \sqrt{3}} & \frac{\sqrt{\frac{2}{3}}}{3} \\
 \frac{1}{2} & \frac{7}{6 \sqrt{3}} & \frac{2 \sqrt{\frac{2}{3}}}{3}
   \\
 \frac{1}{2} & -\frac{7}{6 \sqrt{3}} & -\frac{2
   \sqrt{\frac{2}{3}}}{3} \\
 0 & \frac{5}{3 \sqrt{3}} & -\frac{\sqrt{\frac{2}{3}}}{3} \\
\end{array}
\right),
\end{equation}
and is represented in Figure \ref{FIGK6}.

As in the case of $K_{12}$ it is possible to find a larger group of symmetries of $K_{18}$. We have found a group of order $24$
that is also the automorphism group of the corresponding graph by using the MATHEMATICA command ${\sf  GraphAutomorphismGroup}$.
It follows that the spin sites of $K_{18}$ are not equivalent. They fall into two
subsets of $\{1,2,3,10,11,12\}$ and its complement such that the symmetry group operates transitive only on these subsets but not
on the set of all $18$ spin sites. Nevertheless, it is correct to look for ground states that live in the eigenspace of
${\mathbbm J}$ corresponding to $j_{min}=-2$ and not to consider other gauges in the sense of \cite{S17}. The reason for this is that $j_{min}=-2$  is
equivalent to $E_{min}=-2\times 18 =-36$ as it must be for $12$ corner-sharing triangles in $K_{18}$ with a minimal energy of $-3$ for each triangle.

\section{The Kagome model $K_{12}$ without boundary conditions}\label{sec:K12WB}
\begin{figure}[ht]
  \centering
    \includegraphics[width=1.0\linewidth]{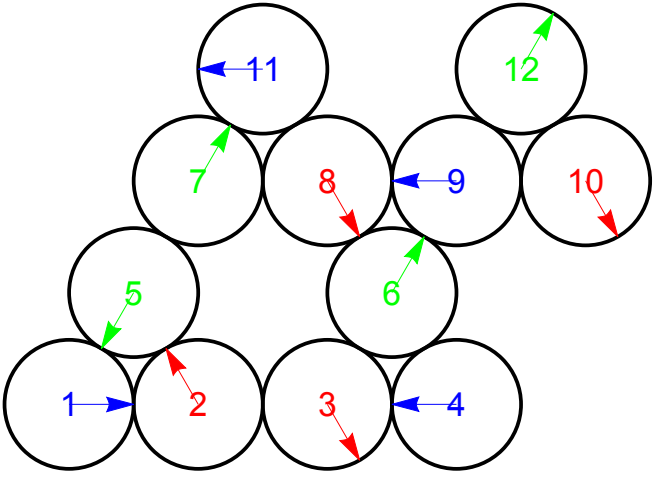}
  \caption
  { Representation of the finite $N=12$ model of the AF Kagome lattice without boundary conditions. A co-planar ground state ${\mathbf c}_{12}$
  with six spin directions forming the vertices of a regular hexagon is indicated by small arrows with the colors red, green, blue.
  All adjacent spin vectors have a scalar product of $-1/2$ except the pairs $(2,3)$ and $(5,7)$ that have $-1$. This gives
  a ground state energy of $-19$ if each bond is counted twice.   If one removes
  all spins of a certain color the remaining graph will be disconnected. Hence there a exist independent rotations of subsets of spins
 about all three spin axes corresponding to the three colors.
   }
  \label{FIGK12A}
\end{figure}

\begin{figure}[ht]
  \centering     \includegraphics[width=1.0\linewidth]{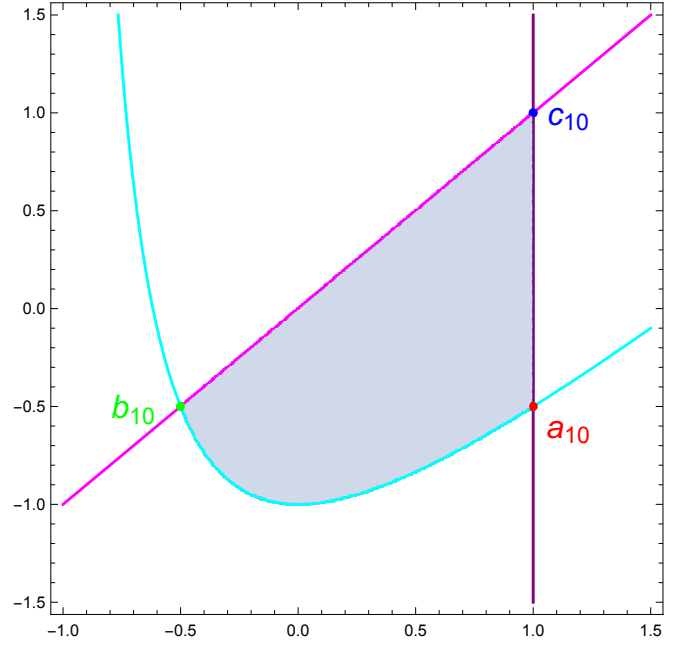}
  \caption
  {The convex set representing ${\mathcal S}_{ADE}$ for the reduced Kagome model $K_{10}$. It is bounded by the lines (\ref{K12WB4a}), (\ref{K12WB4b}),
   and the hyperbola (\ref{K12WB4c}).
  The interior points of ${\mathcal S}_{ADE}$ correspond to $4$-dimensional ground states, and the special extremal points $a_{10},\, b_{10}$ and
  $c_{10}$ to co-planar ground states. The lines between $b_{10},\;c_{10}$ and $a_{10},\;c_{10}$, resp.~, represent a one-dimensional families of $3$-dimensional states generated by independent rotations of subsets of spins, see also Figure \ref{FIGK12C}. The hyperbolic segment joining $a_{10}$ and $b_{10}$
  is another one-dimensional family of $3$-dimensional ground states that does not have such a direct geometric interpretation.
     }
  \label{FIGK12B}
\end{figure}

\begin{figure}[ht]
  \centering
    \includegraphics[width=1.0\linewidth]{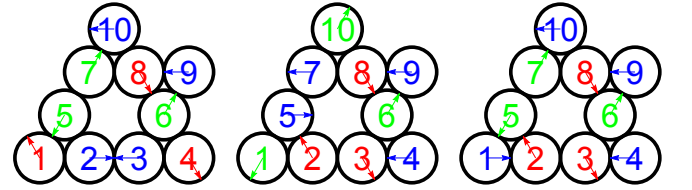}
  \caption
  {From left to right: Representation of the co-planar ground states ${\mathbf a}_{10},\;{\mathbf b}_{10}$ and ${\mathbf c}_{10}$
  of $K_{10}$ corresponding to the three extremal points $a_{10},\, b_{10}$ and   $c_{10}$ of ${\mathcal S}_{ADE}$ in Figure \ref{FIGK12B}.
  One observes that the transition from ${\mathbf b}_{10}$ to ${\mathbf c}_{10}$ is realized by a rotation of the spins $1,5,7,10$ about the
  red spin axis, Analogously,  the transition from ${\mathbf a}_{10}$ to ${\mathbf c}_{10}$ is realized by a rotation of the spins $1,2,3,4$ about the
  green spin axis.
     }
  \label{FIGK12C}
\end{figure}
We consider the Kagome model $K_{12}$, see Figure \ref{FIGK1}, but without boundary conditions and  will denote it  by $K_{12w}$.
Thus the bonds between the spins
$(1, 4), (1, 11), (2, 12), (3, 12), (4, 11), (5, 10), (7, 10)$ are removed and only $17$ bonds remain. This reduces the symmetry group
of $K_{12w}$ to a group of order $4$ generated by two reflections. In general, ground states of this model cannot be extended to the infinite lattice.
Anticipating a co-planar ground state of the form represented in Figure
\ref{FIGK12A} we may conclude that there exist two-parameter families of $3$-dimensional ground states in contrast to $K_{12}$ with boundary
conditions where at most one-parameter families of $3$-dimensional ground states exist.

First we note that the spin with number $9$ separates the graph of $K_{12w}$ into two unconnected parts. In other words, it is possible
to view $K_{12w}$ as a ``fusion" between a spin triangle $(9,10,12)$ and a subsystem $K_{10}$ in the sense of \cite{S17}. It is plausible and has been proven
in \cite{S17} that all ground states of the fusion $K_{12w}$ result from the fusion of two ground states of the spin triangle and $K_{10}$, resp.~.
In particular, to each ground state of $K_{10}$ there corresponds a one-parameter family of ground states of $K_{12w}$ generated by independent
rotation of the $120^\circ$ ground states of the triangle. Hence it suffices to determine the ground states of $K_{10}$ which considerably
simplifies the calculations.

According to previous remarks we now consider the reduced Kagome model $K_{10}$ and re-number the spin sites according to $11 \rightarrow 10$.
In general, every ground state ${\mathbf s}$ of a Heisenberg spin system satisfies the equations
\begin{equation}\label{K12WB1}
\sum_{\nu=1}^N J_{\mu\nu}{\mathbf s}_\nu = - \kappa_\mu\,{\mathbf s}_\mu,\quad \mu=1,\ldots,N
\;,
\end{equation}
where the $\kappa_\mu$  are the Lagrange parameters due to the constraints ${\mathbf s}_\mu\cdot {\mathbf s}_\mu=1,\;\mu=1,\ldots,N$,
and are the same for all ground states of the system, see \cite{S17}.
But in contrast to the Kagome models considered in section \ref{sec:K} in the case of $K_{10}$ the $\kappa_\mu$ are not uniform and we have to modify
the theory outlined in section \ref{sec:D}. We split the $\kappa_\mu$ into the mean value $\overline{\kappa}$ and the deviations $\lambda_\mu$
from the mean value, and add the $\lambda_\mu$ to the diagonal of ${\mathbbm J}$. This yields the ``dressed ${\mathbbm J}$-matrix" denoted by
${\mathbbm J}({\boldsymbol \lambda})$, see \cite{S17} that replaces the undressed ${\mathbbm J}$-matrix.
The further approach is analogous to that outlined in section  \ref{sec:D}.

The Lagrange parameters $\kappa_\mu$ are determined numerically and then rationalized. This works in our case since they are small integers and hence
the $\lambda_\mu$ will be rational numbers with small numerators and denominators. The resulting dressed ${\mathbbm J}$-matrix has the form
\begin{equation}\label{K12WB2}
{\mathbbm J}({\boldsymbol \lambda})=\left(
\begin{array}{cccccccccc}
 -\frac{3}{5} & 1 & 0 & 0 & 1 & 0 & 0 & 0 & 0 & 0 \\
 1 & \frac{2}{5} & 1 & 0 & 1 & 0 & 0 & 0 & 0 & 0 \\
 0 & 1 & \frac{2}{5} & 1 & 0 & 1 & 0 & 0 & 0 & 0 \\
 0 & 0 & 1 & -\frac{3}{5} & 0 & 1 & 0 & 0 & 0 & 0 \\
 1 & 1 & 0 & 0 & \frac{2}{5} & 0 & 1 & 0 & 0 & 0 \\
 0 & 0 & 1 & 1 & 0 & \frac{2}{5} & 0 & 1 & 1 & 0 \\
 0 & 0 & 0 & 0 & 1 & 0 & \frac{2}{5} & 1 & 0 & 1 \\
 0 & 0 & 0 & 0 & 0 & 1 & 1 & \frac{2}{5} & 1 & 1 \\
 0 & 0 & 0 & 0 & 0 & 1 & 0 & 1 & -\frac{3}{5} & 0 \\
 0 & 0 & 0 & 0 & 0 & 0 & 1 & 1 & 0 & -\frac{3}{5} \\
\end{array}
\right),
\end{equation}
and the lowest eigenvalue $j_{min}=-\overline{\kappa}=-\frac{8}{5}$ with $4$-fold degeneracy. The corresponding
eigenspace is spanned by the columns of
\begin{equation}\label{K12WB3}
W=\left(
\begin{array}{cccc}
 -1 & 1 & 0 & -1 \\
 0 & -1 & -1 & 1 \\
 0 & 1 & 1 & -1 \\
 0 & 0 & 0 & 1 \\
 1 & 0 & 1 & 0 \\
 0 & -1 & -1 & 0 \\
 -1 & 0 & -1 & 0 \\
 0 & 0 & 1 & 0 \\
 0 & 1 & 0 & 0 \\
 1 & 0 & 0 & 0 \\
\end{array}
\right).
\end{equation}
The corresponding ADE (\ref{D4}) has a solution $\Delta(x,y)$ depending on two real parameters $x,y$:

\begin{equation}\label{K12WB3a}
\Delta(x,y)=\left(
\begin{array}{cccc}
 1 & x & -\frac{1}{2} & y \\
 x & 1 & -\frac{1}{2} & -x+y+1 \\
 -\frac{1}{2} & -\frac{1}{2} & 1 & x-y-\frac{1}{2} \\
 y & -x+y+1 & x-y-\frac{1}{2} & 1 \\
\end{array}
\right)\;.
\end{equation}

Its determinant is the product of three factors
\begin{eqnarray}\label{K12WB4a}
f_1&=&-1+x\;,\\
\label{K12WB4b}
f_2&=& x-y\;,\\
\label{K12WB4c}
f_3&=&  -1-x+x^2-y-x y \;.
\end{eqnarray}
Hence the convex set ${\mathcal S}_{ADE}$ of solutions $\Delta(x,y)\ge 0$ is represented by the subset of the $(x,y)$-plane bounded by two lines and a hyperbola, see Figure \ref{FIGK12B}. The three extremal points denoted by $a_{10},\; b_{10}$ and $c_{10}$ correspond to the co-planar states represented in
Figure \ref{FIGK12C}. The line segments connecting $c_{10}$ with $b_{10}$ and $a_{10}$, resp.~, correspond to one-parameter families of
$3$-dimensional ground states generated by partial rotations about the red and green spin axis. In contrast, the hyperbolic section
connecting $a_{10}$ with $c_{10}$ corresponds to another one-parameter family of
$3$-dimensional ground states ${\mathbf s}_{10}(x)$ that doesn't have such a direct geometric interpretation. By using the method
described at the end of section \ref{sec:K181} we can
determine the explicit form of this non-rotational family:
\begin{eqnarray}\nonumber
{\mathbf s}_{10}(x)=\hspace{40mm} &&\\
\nonumber&&\\
\label{K12WB5}
\left(
\begin{array}{ccc}
 1 & 0 & 0 \\
 -\frac{1}{2} & \frac{\sqrt{3}}{2} & 0 \\
 \frac{1}{2} & -\frac{\sqrt{3}}{2} & 0 \\
 -x & -\frac{x-1}{\sqrt{3}} & \sqrt{\frac{2}{3}} \sqrt{-2 x^2+x+1} \\
 -\frac{1}{2} & -\frac{\sqrt{3}}{2} & 0 \\
 x-\frac{1}{2} & \frac{2 x+1}{2 \sqrt{3}} & -\sqrt{\frac{2}{3}} \sqrt{-2 x^2+x+1} \\
 \frac{1}{2} & \frac{\sqrt{3}}{2} & 0 \\
 \frac{1}{x+1}-\frac{3}{2} & \frac{x-1}{2 \sqrt{3} (x+1)} & \frac{\sqrt{\frac{2}{3}}
   \sqrt{-2 x^2+x+1}}{x+1} \\
 -x-\frac{1}{x+1}+2 & -\frac{x (x+2)}{\sqrt{3} (x+1)} & \frac{\sqrt{\frac{2}{3}} x
   \sqrt{-2 x^2+x+1}}{x+1} \\
 \frac{x}{x+1} & -\frac{2 x+1}{\sqrt{3} (x+1)} & -\frac{\sqrt{\frac{2}{3}} \sqrt{-2
   x^2+x+1}}{x+1} \\
\end{array}
\right)&&
\end{eqnarray}
for $-\frac{1}{2}< x < 1$.

One should bear in mind that the one-parameter families of ground states of $K_{10}$ just described generate two-parameter families
of ground states of $K_{12w}$ by combination with rotations of ground states of the spin triangle. Especially,
there exist exactly six coplanar ground states of $K_{12w}$ that result from ${\mathbf a}_{10},\;{\mathbf b}_{10}$ and
${\mathbf c}_{10}$ by rotations of the spin triangle ground states with an angle $0^\circ$ and $180^\circ$.

\section{Summary and outlook}\label{sec:SO}
According to \cite{S17} the $O(M)$-equivalence classes of $M$-dimensional ground states of classical Heisenberg systems correspond in a $1:1$ manner
to the points of a convex set ${\mathcal S}_{ADE}$. Physical ground states are typically represented by a subset of its boundary points.
We have completely determined ${\mathcal S}_{ADE}$ and its boundary for the case of the Kagome model $K_{12}$. Here we found,
additional to the well-known co-planar ground states and three rotational families, three non-rotational
families and an isolated $3$-dimensional ground state. For the larger models $K_{15}$ and $K_{18}$ the set ${\mathcal S}_{ADE}$
is too large to be analyzed directly and we confined ourselves to certain subsets of symmetric ground states.
Also with this restriction we could identify some non-rotational families and isolated ground states.
A complete enumeration of all physical ground states was also possible for the Kagome model $K_{12w}$ without boundary conditions
and accordingly less symmetry.

These case studies raise a couple of physical, computational and mathematical questions.
First, one may ask whether the new kinds of ground states of the Kagome lattice
are relevant for its low-temperature behavior, e.~g., concerning the specific heat or correlation functions. Numerical evidence suggests
that this is not the case, see \cite{CHS92} and \cite{Z08}, but one would like to understand the reason.

Second, if it becomes difficult to analyze ${\mathcal S}_{ADE}$ for larger number of spins $N$ one would rather try to analyze its
faces in order to find physical ground states. One way to do this is to find a factorization of $\det(\Delta)$, see (\ref{D3}),
but this will not be always as simple as in the case of (\ref{K12WB4a})--(\ref{K12WB4c}). How should one find these factorizations
in the general case? Another way to explore the boundary of ${\mathcal S}_{ADE}$ is to look for its singular part, defined by the
simultaneous vanishing of $\det(\Delta)$ and its gradient. We found that typically the rank of $\Delta$ is reduced by more than one if crossing
the boundary of ${\mathcal S}_{ADE}$ at a singular point, but a general mathematical theory covering these effects would be desirable.

\section*{Acknowledgment}
I would like to thank the members of the groups of Johannes Richter (Magdeburg) and J\"urgen Schnack (Bielefeld) for discussions
on the subject of this paper. Further, I am indebted to Thomas Bilitewski for communicating detailed numerical results related to \cite{BZM17}.

\end{document}